\documentclass[sigconf]{acmart}
\usepackage{multirow}
\usepackage{float}
\usepackage{subfig}
\usepackage{enumitem}
\usepackage{algorithmic}
\newtheorem{assumption}{Assumption}

\usepackage[ruled, linesnumbered]{algorithm2e}

\newcommand{\fullname}{\textit{Social Graph Invariant Learning(SGIL)}}
\newcommand{\shortname}{\textit{SGIL}}

\AtBeginDocument{%
  \providecommand\BibTeX{{%
    \normalfont B\kern-0.5em{\scshape i\kern-0.25em b}\kern-0.8em\TeX}}}

\setcopyright{acmlicensed}

\copyrightyear{2025}
\acmYear{2025}
\setcopyright{acmlicensed}\acmConference[SIGIR '25]{Proceedings of the 48th
International ACM SIGIR Conference on Research and Development in
Information Retrieval}{July 13--18, 2025}{Padua, Italy}
\acmBooktitle{Proceedings of the 48th International ACM SIGIR Conference on
Research and Development in Information Retrieval (SIGIR '25), July 13--18,
2025, Padua, Italy}
\acmDOI{10.1145/3726302.3730013}
\acmISBN{979-8-4007-1592-1/2025/07}

\begin{document}

\title{Invariance Matters: Empowering Social Recommendation \\
via Graph Invariant Learning}

\author{Yonghui Yang}
\orcid{0000-0002-7601-6004}
\affiliation{
\department{Key Laboratory of Knowledge Engineering with Big Data,}
\institution{Hefei University of Technology}
\city{Hefei}
\country{China}
}
\email{yyh.hfut@gmail.com}

\author{Le Wu}
\orcid{0000-0003-4556-0581}
\authornotemark[1]
\affiliation{
\department{Key Laboratory of Knowledge Engineering with Big Data,}
\institution{Hefei University of Technology}
\department{Institute of Dataspace,}
\institution{Hefei Comprehensive National Science Center}
\city{Hefei}
\country{China}
}
\email{lewu.ustc@gmail.com}
\thanks{* Corresponding author}

\author{Yuxin Liao}
\orcid{0009-0001-8421-1061}
\affiliation{
\department{Key Laboratory of Knowledge Engineering with Big Data,}
\institution{Hefei University of Technology}
\city{Hefei}
\country{China}
}
\email{yuxinliao314@gmail.com}

\author{Zhuangzhuang He}
\orcid{0000-0001-6608-2940}
\affiliation{
\department{Key Laboratory of Knowledge Engineering with Big Data,}
\institution{Hefei University of Technology}
\city{Hefei}
\country{China}
}
\email{hyicheng223@gmail.com}

\author{Pengyang Shao}
\orcid{0000-0003-2838-1987}
\affiliation{
\department{Key Laboratory of Knowledge Engineering with Big Data,}
\institution{Hefei University of Technology}
\city{Hefei}
\country{China}
}
\email{shaopymark@gmail.com}

\author{Richang Hong}
\orcid{0000-0001-5461-3986}
\affiliation{
\department{Key Laboratory of Knowledge Engineering with Big Data,}
\institution{Hefei University of Technology}
\department{Institute of Dataspace,}
\institution{Hefei Comprehensive National Science Center}
\city{Hefei}
\country{China}
}
\email{hongrc.hfut@gmail.com}

\author{Meng Wang}
\orcid{0000-0002-3094-7735}
\affiliation{
\department{Key Laboratory of Knowledge Engineering with Big Data,}
\institution{Hefei University of Technology}
\city{Hefei}
\country{China}
}
\email{eric.mengwang@gmail.com}

\renewcommand{\shortauthors}{Yonghui Yang et al.}

\begin{abstract}
Graph-based social recommender systems have demonstrated great potential in alleviating data sparsity by leveraging high-order user influence embedded in social networks. However, most existing methods rely heavily on the observed social graph, which is often noisy and includes spurious or task-irrelevant connections that can mislead user preference learning. Identifying and removing these noisy relations is crucial but challenging due to the lack of ground-truth annotations. In this paper, we approach the social denoising problem from the perspective of graph invariant learning and propose a novel approach, \fullname. Specifically, \shortname~aims to uncover stable user preferences within the input social graph, thereby enhancing the robustness of graph-based social recommendation systems. To achieve this goal, \shortname~first simulates multiple noisy social environments through graph generators. It then seeks to learn environment-invariant user preferences by minimizing invariant risk across these environments. To further promote diversity in the generated social environments, we employ an adversarial training strategy to simulate more potential social noisy distributions. Extensive experimental results demonstrate the effectiveness of the proposed \shortname. The code is available at https://github.com/yimutianyang/SIGIR2025-SGIL.
\end{abstract}



\maketitle

\section{Introduction}
\begin{figure*}[th]
    \centering
    \includegraphics[width=165mm]{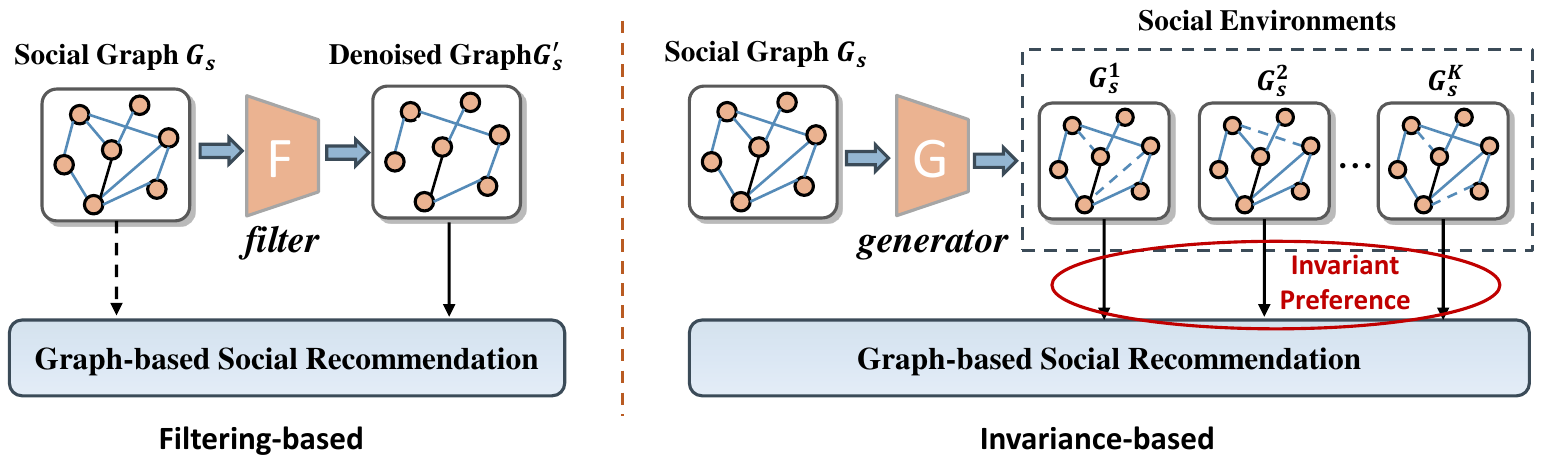}
    \caption{Illustration of the existing filtering-based social denoising methods and our proposed invariance-based method. Instead of filtering the noisy part of the input social graph, we encourage learning the invariant yet sufficient user preferences across environments.}
    \label{fig:intro}
\end{figure*}

With the proliferation of social platforms, social recommendation has emerged as a pivotal technique for delivering personalized suggestions to users, harnessing both user-item interaction data and user-user social relationships~\cite{tang2013social}. The underlying premise is that users who are socially connected are more likely to exhibit similar preference patterns~\cite{mcpherson2001birds}. Consequently, much of the prior research has focused on how to effectively leverage the potential of social networks to enhance the learning of user preferences~\cite{ma2008sorec, konstas2009social}. Early work in this domain predominantly relied on first-order social connections to improve recommendation quality. For instance, user preferences were enhanced through social regularization~\cite{jamali2010matrix} or by incorporating trust relationships into user preference factorization~\cite{guo2015trustsvd}. These methods paved the way for more sophisticated integration of social data.

In recent years, graph-based social recommendation methods have gained prominence due to their remarkable success in improving recommendation accuracy, particularly in addressing the challenge of data sparsity within user behaviors~\cite{LightGCN, DiffNet, hu2024mgdcf}. By leveraging the powerful representational capabilities of Graph Neural Networks (GNNs), these methods encode user preferences while simultaneously capturing higher-order social influences and collaborative signals~\cite{wu2020diffnet++, DiffNet}. This has made graph-based approaches a compelling choice for incorporating social information into recommendation systems, enabling them to better model the complex relationships between users and their social networks.

However, despite the apparent benefits of aggregating all available information from social networks, many graph-based social recommendation methods fail to account for the inevitable noise inherent in these networks. Social networks are often fraught with task-irrelevant relationships that may obscure or distort the true preferences of users. Previous works~\cite{WWW2023robust, jiang2024challenging, yang2024graph} have empirically and statistically demonstrated the pervasiveness of social noise. For example, ~\cite{jiang2024challenging} defines the concept of preference-aware social homophily ratio and observes that social networks typically exhibit low homophily, meaning that social connections do not always align with user preferences. Furthermore, \cite{yang2024graph} investigates the direct use of original social networks in graph-based recommenders, revealing that such approaches result in only incremental improvements, or in some cases, a decline in performance. This highlights the importance of addressing the noise present in social data for effective recommendation learning.

Nevertheless, the task of identifying and eliminating these noisy social relations remains a substantial challenge. Current approaches typically follow a two-step paradigm: 1) filtering the social structure based on user preference signals, and 2) optimizing this filtering process using specific learning objectives, such as co-optimization~\cite{WWW2023robust, wsdm2024madm}, contrastive learning~\cite{jiang2024challenging}, and graph bottleneck learning~\cite{yang2024graph}. Although these methods have made notable strides in denoising social data, they are inherently limited by the lack of reliable mechanisms to identify and verify noise, as there are no ground truth labels available to assess the accuracy of the filtered relationships. This uncertainty raises an important question: \textbf{\emph{Given the difficulty in exhaustively removing noisy social connections, can we still learn stable and sufficient user preferences, even in the presence of noisy environments?}}

To achieve this goal, we adopt the principle of invariant learning to address the social noise issue. Invariant learning~\cite{IRM, vRex} posits that the input data contains both invariant features (stable) and spurious features (variable across different environments). The focus of invariant learning is to identify and utilize only the invariant features, thereby enhancing the model's robustness. As illustrated in Figure \ref{fig:intro}, we compare existing filtering-based social denoising methods with our proposed invariance-based approach. Drawing inspiration from this principle, we hypothesize that a task-relevant structure exists within the social graph that is both sufficient and invariant for recommendation tasks. Although this invariant structure is not directly accessible, we propose narrowing the bounds of invariance by exploring diverse noisy environments. Specifically, for a given input social graph, we first simulate multiple social environments (which embody different types of noise), and then optimize the learning objectives to encourage invariance across these environments. This formulation presents two key challenges: i) How can we simulate diverse noisy environments based on the input social networks and interactions? ii) How can we achieve invariant user preference learning across different noisy environments?

In this work, we propose a novel \fullname~, to tackle the above challenges. First, we design preference-guided environment generators that simulate multiple noisy environments, thereby approximating real-world scenarios where social noise is prevalent. Next, we capture invariant user preferences by leveraging an invariance-based regularizer. Specifically, we derive the optimization objective for each environment, taking into account the characteristics of the recommendation task, and incorporate an invariance penalty term to enforce the stability of user preferences across environments. To further ensure the diversity of the simulated noisy environments, we employ an adversarial training strategy. This strategy challenges the model iteratively, forcing it to perform well in progressively more difficult and varied environments, which ultimately improves its generalization ability.
The main contributions of this work are summarized as follows: 
\begin{itemize}[leftmargin=1.2em]
    \item We propose a novel social denoising approach \shortname~for robust recommendation from the standpoint of invariant learning. Unlike traditional methods that attempt to remove noisy data directly, \shortname~focuses on learning invariant user preferences across diverse environments, ensuring robustness despite noise. 
    \item We derive an optimization objective tailored to each social environment and design an invariance-based optimization to learn stable user preferences. Additionally, we incorporate an adversarial training strategy to guarantee the diversity of the generated environments, further enhancing the model's robustness. 
    \item We conduct extensive experiments on three benchmark datasets. Empirical studies, including both real and semi-synthetic datasets, demonstrate the effectiveness of the proposed \shortname~method in improving social recommendation performance and robustness. 
    \end{itemize}

\section{Related Works}
\subsection{Graph-based Social Recommendation}
Social recommendations have been widely deployed in online services, which leverage social networks as auxiliary information to capture users' behavior patterns~\cite{marsden1993network, tang2013social}. Following the social homophily~\cite{mcpherson2001birds} and social influence theory~\cite{marsden1993network}, social recommendations are devoted to characterizing social relation effects on user preferences, including trust-based preference learning~\cite{guo2015trustsvd}, social regularization~\cite{jamali2010matrix, ma2011recommender}, and social influence modeling~\cite{DiffNet}. Early works mainly exploit shallow social relations to enhance recommendation, ignoring the complex high-order social structure influences. 
 
Users' interactions can be naturally formulated in graph forms~\cite{LightGCN, yang2023generative}, many graph-based social recommendations have been proposed~\cite{DiffNet, SEPT, HGSR}. 
Particularly, graph-based social recommendations achieved an impressive process by borrowing the power of representation ability of GNNs.The core premise of this work is that incorporating high-order social neighbors and interacted items into user preference learning enriches the available information, thereby addressing the issue of data sparsity. Current studies mainly focus on modeling social structure in Euclidean space~\cite{wu2020diffnet++, SEPT}, few attempts also explore the potential of Hyperbolic space learning~\cite{TOIS2021hypersorec, HGSR}. However, despite the effectiveness of social graph modeling, current research often overlooks the challenge of social noise. Despite the effectiveness of social graph modeling, current studies rarely notice the social noise problem. Directly using the observed social graph may lead to sub-optimal results. In this work, we propose a novel social denoising method to alleviate the effect of task-irrelevant social structures in recommendations.

\subsection{Recommendation Denoising}
Recommender systems are usually built on clean label assumption, but it's hard to guarantee in real-world datasets. Therefore, recommendation denoising methods are designed to collaborative filtering~\cite{wang2021denoising, yang2021enhanced, he2024double}, multimodal recommendation~\cite{ma2024multimodal, yang2025less, hu2025modality}, and social recommendation~\cite{jiang2024challenging, wsdm2024madm, yang2024graph} scenarios.
Specifically, researchers design social denoising methods to reduce the effect of redundant social structures~\cite{TKDE2020enhancing, WWW2023robust, jiang2024challenging, yang2024graph}. Among these, ESRF leverages the adversarial training technique to generate potential social relations and filter unstable part~\cite{TKDE2020enhancing}. GDSMR~\cite{WWW2023robust} proposes a distilled social graph based on progressive preference-guided social denoising. ShaRe~\cite{jiang2024challenging} designs a social graph rewriting method with social-enhanced contrastive learning optimization. GBSR~\cite{yang2024graph} introduces the information bottleneck principle to guide the social denoising process, which can preserve the minimal yet sufficient social structure. In summary, current solutions aim to filter out the noisy part of social networks, but identifying the noisy part is challenging due to a lack of labels. Instead of removing the noise part directly, we focus on learning stable user preferences across diverse environments~(which are full of various noise). 
If the model can survive in various noisy environments, it indeed captures the invariant social structure for recommendation tasks. 
Thus, we introduce the invariant learning principle to tackle the social noise problem.

\subsection{Invariant Learning and Applications}
Invariant learning has emerged as a promising approach to address the challenges posed by Out-Of-Distribution (OOD) generalization in machine learning~\cite{IRM, vRex, EIIL, yue2025learning}. The core idea behind invariant learning is to uncover stable feature representations that remain robust under varying environmental conditions, thereby enhancing the reliability of label prediction models. This framework is based on the assumption that the data contains both spurious and invariant features, where spurious features vary unpredictably across different environments, while invariant features remain stable and consistent. The primary objective of invariant learning methods is to optimize models using Invariant Risk Minimization (IRM) by introducing an invariance-based penalty during optimization, which discourages reliance on spurious features and encourages the model to learn stable, generalizable patterns. 
Invariant learning has been successfully applied across a wide range of machine learning tasks, including image classification~\cite{IRM, EIIL}, graph learning~\cite{li2022learning, chen2022learning, ICLR2022discovering}, and recommender systems~\cite{MM2022invariant, wang2022invariant, WWW2023invariant, WWW2024unleashing, SIGIR2024multimodality}. Especially for recommender systems, ~\cite{MM2022invariant, SIGIR2024multimodality} learn invariant multimedia representation to improve the robustness of the general and cold-start recommendation scenarios. Besides, ~\cite{wang2022invariant, WWW2023invariant} leverage the invariant learning to address popularity debiasing, and ~\cite{WWW2024unleashing} captures the invariant KG subgraph to enhance recommendations.
In this work, we introduce the invariant learning principle to the social recommendation, intending to capture stable user preferences that persist across diverse social environments.

\section{Preliminary}
\subsection{Graph-based Social Recommendation}
In the fundamental social recommender, there are two kinds of entities: a userset $U$~($|U|=M$) and an itemset $V$~($|V|=N$). Users present two kinds of behaviors, user-user social behaviors~(e.g., follow, favor, group) and user-item interaction behaviors~(e.g., click, view, purchase). We use the matrix $\mathbf{S}\in \mathbb{R}^{M \times M}$ to describe the user-user social relation matrix, where each element $\mathbf{s}_{ab}=1$ if user $b$ has social behavior with user $a$, otherwise $\mathbf{s}_{ab}=0$. Meanwhile, we use the matrix $\mathbf{R} \in \mathbb{R}^{M \times N}$ to describe the user-item interaction behaviors, where each element $\mathbf{r}_{ai}=1$ if user $a$ interacted with item $i$. 
Given the userset $U$, itemset $V$, user-user social relations $\mathbf{S}$ and user-item interactions $\mathbf{R}$, we construct the user-item interaction graph $\mathcal{G}_r=\{U \cup V, \mathcal{E}_r=\{(a,i)|\mathbf{r}_{ai}=1\}\}$ and the user-user social graph $\mathcal{G}_s=\{U, \mathcal{E}_s=\{(a,b)|\mathbf{s}_{ab}=1\}\}$. Graph-based social recommendation aims to infer the unknown preference of user $a$ to item $i$:
$\hat{r}_{ai} = f_{\phi}(a,i,\mathcal{G}_r,\mathcal{G}_s))$, where $f_{\phi}$ denotes the GNN formulation with parameters $\phi$. The traditional learning paradigm focuses on Empirical Risk Minimzation~(ERM):
\begin{flalign}
    Min: \mathbb{E}_{(a,i,\mathbf{r}_{ai})\sim \mathcal{P}} \mathcal{L}(r_{ai}; f_{\phi}(a,i,\mathcal{G}_r,\mathcal{G}_s)),
\end{flalign}

\noindent where $\mathcal{P}$ denotes the distribution of training data. However, the observed social networks are inevitably noisy with task-irrelevant relations, and directly using them will lead to sub-optimal results~\cite{jiang2024challenging, yang2024graph}. Therefore, tackling the social noise issue is an urgent need to enhance recommendations.

\subsection{Graph Invariant Learning}
Invariant learning is the classic branch of Out-Of-Distribution ~(OOD) generalization methods, tackling the spurious correlations in machine learning tasks~\cite{IRM, EIIL, vRex}. Graph invariant learning focuses on Graph OOD generalization, targeting the pursuit of invariant subgraph across diverse noisy environments~\cite{wu2022handling, li2022learning, chen2022learning, xia2024learning}. Typically, graph invariant learning methods have the following assumption:

\begin{assumption}
    Given the input graph $\mathcal{G}$, there exists a subgraph $\mathcal{G}^I \in \mathcal{G}$ satisfy the invariance assumption: (1) Sufficiency condition: $y=f(\mathcal{G}^I)+\epsilon$, where $y$ is the prediction target and $\epsilon$ is an independent noise. (2) Invariance condition: for any two environments $e, e' \in \mathcal{E}$, $p(y|\mathcal{G}^I, e)=p(y|\mathcal{G}^I, e')$, where $\mathcal{E}$ denotes all potential environments. 
\end{assumption}

This assumption means that the invariant subgraph involves sufficient information from the input graph, which can build a robust prediction model for unknown distributions. To achieve this goal, an Invariant Risk Minimization~(IRM) based optimization has been proposed~\cite{IRM, vRex}, which is defined as follows:
\begin{flalign}
    Min: \mathbb{E}_{e\in \mathcal{E}} \mathcal{L}_e + \beta Var(\{\mathcal{L}_e, e\in \mathcal{E}\}),
\end{flalign}

\noindent where the first term focuses on the empirical risk minimization in each environment~(sufficiency condition), and the second term encourages the invariant risk minimization~(invariance condition). The parameter $\beta$ is set to find the trade-off between sufficiency and invariance. 
Inspired by graph invariant learning, we propose a novel \fullname~approach to tackle the social noise issue. Imaging that there exists an invariant social subgraph for recommendation, while other redundant relations construct various noisy environments. Following the invariance assumption, can we learn the invariant social subgraph to facilitate the recommendation task? Let's begin with the technical details.

\begin{figure*}[t]
    \centering
    \includegraphics[width=170mm]{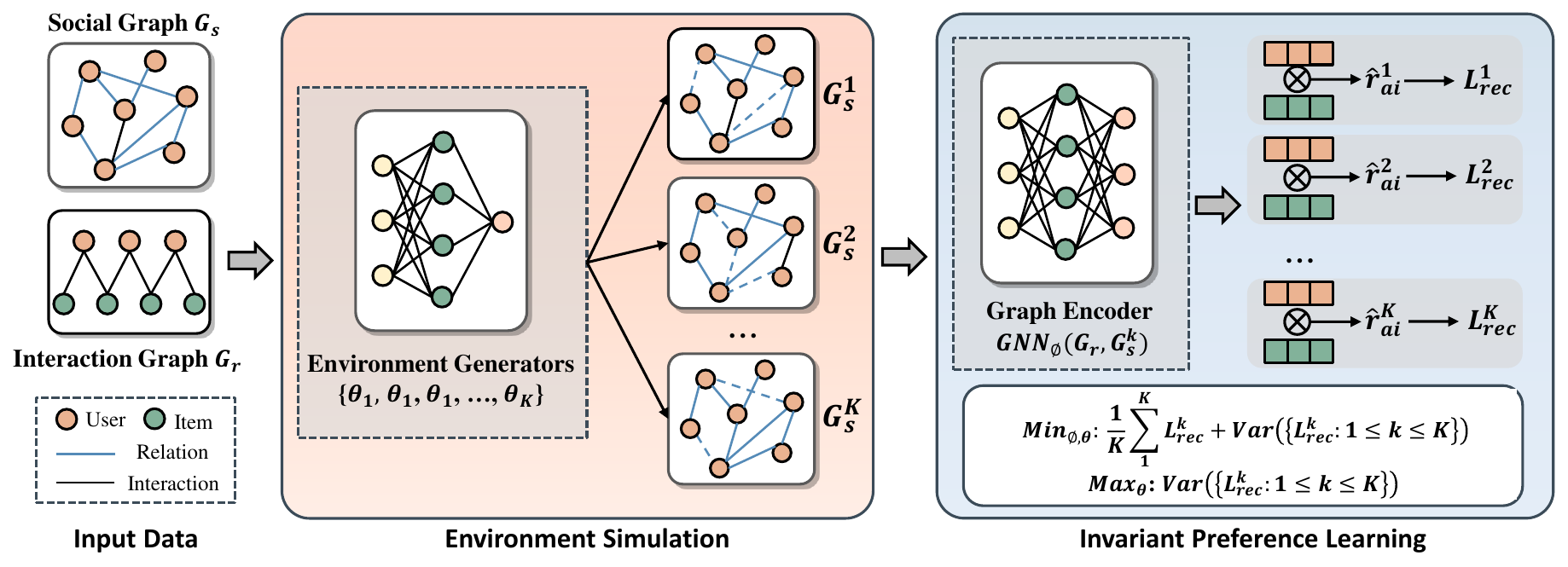}
    \vspace{-0.3cm}
    \caption{Overview of our proposed \shortname~framework, consists of two elaborated modules. (1) Environment Simulation: aiming to generate multiple social environments; (2) Invariant Preference Learning: targeting to achieve the invariant user preference across environments.}
    \label{fig:framework}
\end{figure*}

\section{Methodology}
As illustrated in Figure 2, we present the overall framework of our proposed \fullname~approach. Essentially, \shortname~aims to learn stable yet sufficient user preferences across diverse social environments. To achieve this goal, \shortname~elaborates two modules: Environment Simulation, and Invariant Preference Learning. Specifically, the environment simulation module aims to generate multiple social environments, and the invariant preference learning module targets to achieve the invariant user preference across environments. Next, we introduce each module in detail.

\subsection{Environment Simulation}
Given the user-user social graph $\mathcal{G}_s$ and user-item interaction graph $\mathcal{G}_s$ as input, we first simulate diverse noisy environments. Here, we define K social environments $\mathcal{E}=\{e_k: 1 \leq k \leq K
\}$. Consequently, we define K independent environment generators $\{g_{\theta}^k\}_{k=1}^K$ with parameters $\{\theta^1,\theta^2,...,\theta^K\}$. We take $k^{th}$ social environment as an example to illustrate how to construct it.
Formally, constructing $k^{th}$ social environment equals extracting $k^{th}$ social subgraph $\mathcal{G}_s^k$, each edge $(a,b)$ will be preserved with a probability $\rho_{ab}$. Thus, we have the extracted subgraph structure $\mathbf{S}^k$:
\begin{flalign}
    \label{eq: subgraph extraction}
    \mathbf{S}^k = g_{\theta}^k(U,\mathcal{G}_r,\mathcal{G}_s) = \{\mathbf{s}_{ab} \odot \rho_{ab}^k\},
\end{flalign}

\noindent where $\rho_{ab}^k \sim Bern(w_{ab}^k)+\epsilon$ denotes that each edge $<u_a,u_b>$ will be dropped with the probability $1-w_{ab}^k+\epsilon$. 

Following the previous work~\cite{yang2024graph}, we also add the bias term $\epsilon=0.5$ to guarantee that the confidences of observed social relations are higher than those non-observed. Following the previous studies~\cite{yang2024graph}, we feed users' collaborative singles into parameter $w_{ab}^k$ calculation.
Let $\mathbf{E}^U \in \mathbb{R}^{M \times d}$ and $\mathbf{E}^I \in \mathbb{R}^{N \times d}$ denote user and item preference representations, which learned from the input data $\{\mathcal{G}_r, \mathcal{G}_s\}$. We parameterize the distribution parameter $w_{ab}^k$ as follows:
\begin{flalign} \label{eq: drop probability}
    w_{ab}^k = (g_{\theta}^k(\mathbf{e}_a, \mathbf{e}_b)),
\end{flalign}

\noindent where $\mathbf{e}_a$ and $\mathbf{e}_b$ denote user $a's$ and user $b's$ preference representations, respectively. Besides, we implement the generator $g_{\theta}^k(\cdot)$ through two-layer MLPs with parameters $\theta^k$. After the above learning process, we can obtain $k^{th}$ social subgraph $\mathcal{S}^k$. However, the discrete Bernoulli distribution is not differentiable. Gumbel-Softmax operation is the popular solution to relax the learning process:
\begin{flalign} \label{eq: drop reparameter}
    Bern(w_{ab})=sigmoid({log(\delta/(1-\delta)+w_{ab})}/t),
\end{flalign}

\noindent where $\delta \sim U(0,1)$, and $t \in \mathbb{R^{+}}$ is the temperature parameter~(we set t=0.2 in our experiments). 

After the re-parameterization operation, the discrete Bernoulli distribution is transferred to a differentiable function. Repeating $K$ times subgraph extraction $\{\mathbf{S}^1, \mathbf{S}^2,...,\mathbf{S}^K\}$, we can simulate $K$ social environments $\{\mathcal{G}_s^1, \mathcal{G}_s^2, ..., \mathcal{G}_s^K\}$ .

\subsection{Invariant Preference Learning}
Given the simulated social environments, \shortname~then learns the invariant preference across diverse environments. The invariant learning module consists of three components: graph-based preference learning, empirical risk minimization, and overall optimization with invariance. 

\subsubsection{Graph-based Preference Learning.}
Due to the advanced representation ability of GNNs, here we use GNNs as encoders to capture users' preferences. Specifically, we employ LightGCN-S~\cite{LightGCN, yang2024graph} as the encoder to obtain user preference embeddings. We first formulate the user-item interactions and the user-user social relations in $k^{th}$ environment as a heterogeneous graph
$\mathcal{G}^k=\{U \cup V, \mathbf{A}^k\}$, where $U \cup V$ denotes the set of nodes, and $\mathbf{A}^k$ is the adjacent matrix defined as follows:
\begin{flalign}\label{eq:adj_matrix}
\mathbf{A}^k=\left[\begin{array}{cc}
\mathbf{S^k} & \mathbf{R}\\
\mathbf{R}^T & \mathbf{0}^{N\times N}
\end{array}\right].
\end{flalign}

\noindent Let $\mathbf{P}^0 \in \mathbb{R}^{M \times d}$ and $\mathbf{Q}^0$ denote the initialized user and item embeddings, where $d$ is the embedding dimension. Then, we iteratively update node embeddings through graph convolutions:
\begin{flalign}\label{eq: GCN form}
\left[\begin{array}{c}
\mathbf{P}^{l+1} \\
\mathbf{Q}^{l+1}\end{array}\right]
=\mathbf{D}_k^{-\frac{1}2}\mathbf{A}^k\mathbf{D}_k^{-\frac{1}{2}} \times
\left[\begin{array}{c}
\mathbf{P}^{l} \\
\mathbf{Q}^{l}\end{array}\right],
\end{flalign}

\noindent where $\mathbf{D}_k$ is the degree matrix of graph $\mathcal{G}^k$. When stacking $L$ graph convolution layers, we have the final node embeddings through the avg-pooling operation:
\begin{flalign}\label{eq:readout}
\mathbf{P} =\frac{1}{L+1}\sum_{l=0}^L\mathbf{P}^l, \mathbf{Q} =\frac{1}{L+1}\sum_{l=0}^L\mathbf{Q}^l.
\end{flalign}

\noindent After obtaining the learned node representations in the $k^{th}$ social environment, we can infer the propensity that user $a$ will interacts with item $i$ by an inner product: $\hat{r}_{ai}=<p_a^k, q_i^k>$. All the above process are summarized as $\hat{r}_{ai}^k=f_{\phi}^k(a,i,\mathcal{G}_r, \mathcal{G}_s^k)$, where $\phi=\{\mathbf{P}^0, \mathbf{Q}^0\}$ denote the initialized embedding parameters. 

\subsubsection{Empirical Risk Minimization.}
Following the above use preference learning procedure, we obtain K predictions across all environments. Next, we implement the sufficiency condition in Assumption 1. Given the predictions in $k^{th}$ environment, the empirical risk loss is defined as follows:
\begin{flalign}
    \mathcal{L}_{rec}^k = \sum_{a\in U, i\in V} \mathcal{L}(\sigma(f_{\phi}^k(a,i,\mathcal{G}_r,\mathcal{G}_s^k)), \mathbf{r}_{ai}).
\end{flalign}

\noindent Although this point-wise loss satisfies the sufficiency condition, in practice, we find that it's hard to guarantee the invariance condition. The reason is that the user's interaction behaviors are not IID data~\cite{cao2016non}, directly aligning the user's preference to a specific item under various environments can not guarantee invariant user preference. In other words, we must align each user preference distribution across all environments. Therefore, we rewrite the empirical risk loss as:
\begin{flalign}
    \label{eq: distribution loss}
    \mathcal{L}_{rec}^k = \sum_{a\in U} \mathcal{L}(\sigma(f_{\phi}^k(a,V,\mathcal{G}_r,\mathcal{G}_s^k)), \mathbf{R}_{a}).
\end{flalign}

\noindent Next, we introduce how to optimize the above loss function. Let $V_a(|V_a|=d_a)$ denote each user $a's$ interacted items, then we have the prior information: $p(r=1)=\frac{d_a}{N}$ and $p(r=0)=\frac{N-d_a}{N}$. Next, we analyze the interaction probability as follows:
\begin{equation}
    \begin{aligned}
        p(r=1|a,i) &= \frac{p(r=1)p(a,i|r=1)}{\sum_{r\in \{0,1\}}p(r)p(a,i|r)} \\
    &= \frac{\frac{d_a}{N} p(a,i|r=1)}{\frac{d_a}{N} p(a,i|r=1)+\frac{N-d_a}{N} p(a,i|r=0)} \\
    &= \frac{p(a,i|r=1)}{p(a,i|r=1)+\frac{N-d_a}{d_a}p(a,i|r=0)} \\
    &= \frac{p(a,i|r=1)}{p(a,i|r=1)+C p(a,i|r=0)},
    \end{aligned}
    \label{eq: nce}
\end{equation}

\noindent where $C=\frac{N-d_a}{d_a}$ is the negative sample ratio. In $k^{th}$ social environment, we infer the probability $p(r=1|a,i)=\delta(f_{\phi}^k(a,i,\mathcal{G}_r, \mathcal{G}_s^k))$, where $\delta(\cdot)$ is the sigmoid function. For the sake of clarity, we use $f_{\phi}^k(a,i)$ to replace $f_{\phi}^k(a,i,\mathcal{G}_r, \mathcal{G}_s^k))$, then we define the density ratio as:
\begin{flalign}
    \frac{p(a,i|r=1)}{p(a,i|r=0)} &= C \frac{p(r=1|a,i)}{1-p(r=1|a,i)} = C exp(f_{\phi}^k(a,i)).
\end{flalign}

\noindent In practice, directly optimizing the scores of full items is consuming, so we employ the popular batch training strategy. Previous works demonstrate that the more negative samples, the better the conditional probability estimation $p(r=1|a,i)$. For batch training data with $\mathcal{I}_b$ items,
supposing there are one positive sample $(a,i)$ and $|\mathcal{I}_b|-1$ negative samples $(a,j)$ in the training data, 
the goal is to maximize the following probability:

\begin{equation}
\begin{aligned}
    P &= \frac{p(a,i|r=1)\prod_{j!=i, j\in \mathcal{I}_b}p(a,j|r=0)}{\sum_{i \in \mathcal{I}_b}p(a,j|r=1)\prod_{w!=j, w\in \mathcal{I}_b}{p(a,w|r=0)}} \\
    &= \frac{p(a,i|r=1)/p(a,i|r=0)}{\sum_{j\in \mathcal{I}_b} p(a,j|r=1)/p(a,j|r=0)} \\
    &= \frac{exp(f_{\phi}^k(a,i))}{\sum_{j \in \mathcal{I}_b} exp(f_{\phi}^k(a,j))}.
\end{aligned}
\end{equation}

\noindent  We use the popular scaled cosine similarity~\cite{wu2024effectiveness} to implement $f_{\phi}^k(a,i)$, thus we can obtain the final optimization objective of ERM in each environment:
\begin{flalign}
    \label{eq: SSL}
    \mathcal{L}_{rec}^k = -\frac{1}{|\mathcal{D}|}\sum_{(a,i) \in \mathcal{D}}\frac{exp(<p_a^k, q_i>/\tau)}{\sum_{j \in \mathcal{I}_b}exp(<p_a^k,q_j^k>/\tau)}.
\end{flalign}

\subsubsection{Overall Optimization with Invariance.} 
Given the above optimization objective of ERM in each social environment, we then combine the sufficiency condition and invariance condition for optimization. Specifically, we achieve the overall optimization objective with a variance-based regularizer:
\begin{flalign}
\label{eq: overall loss}
    Min: \frac{1}{K} \sum_{k=1}^K \mathcal{L}_{rec}^k + \beta Var(\{\mathcal{L}_{rec}^k: 1 \leq k \leq K\}),
\end{flalign}

\noindent where the first term focuses on the empirical risk minimization in each social environment~(sufficiency condition), and the second term encourages the invariant risk minimization~(invariance condition), and $\beta$ is the trade-off parameter. To enhance the robustness of \shortname~in more diverse noisy environments, we further conduct the environment exploration based on an adversarial training strategy. Specifically, we iteratively optimize the following two objectives:
\begin{flalign}
\label{eq: optimization}
    \phi^*, \theta^* &= \mathop{\arg\min}\limits_{\phi, \theta} \frac{1}{K} \sum_{k=1}^K \mathcal{L}_{rec}^k + \beta Var(\{\mathcal{L}_{rec}^k: 1 \leq k \leq K\}) \\
    \theta^* &= \mathop{\arg\max}\limits_{\theta} Var(\{\mathcal{L}_{rec}^k: 1 \leq k \leq K\}).
\end{flalign}

\noindent Considering the instability of adversarial training, we select a cross-batch training strategy. We will conduct the exploration process every $T$ batch. After cross-environment optimization, we can obtain a stable user preference for the recommendation. 
In the inference stage, we first compute the mean representations of users and items across $K$ social environments:
\begin{flalign}
    \mathbf{U} = \frac{1}{K} \sum_{k=1}^K \mathbf{P}^k, \mathbf{V} = \frac{1}{K} \sum_{k=1}^K \mathbf{Q}^k.
\end{flalign}

\noindent Then, we infer the interaction score based on the inner product: $\hat{r}_{ai}=<\mathbf{u}_a, \mathbf{v}_i>$. 
We summarize the learning process of \shortname~ as Algorithm 1.

\begin{algorithm}[t]
\caption{The algorithm of \shortname~}
\begin{algorithmic}[1]
\STATE \textbf{Input:} Userset $U$, Itemset $V$, user-item interactions $\mathbf{R}$, user-user social relations $\mathbf{S}$, the number of social environments $K$, and observation bias $\epsilon$;
\STATE \textbf{Output:} Optimal graph-based social recommender $\mathcal{G}^{*}_{\theta, \phi}(\cdot)$, where $\theta$ denote MLPs used in environment generators and $\phi=\{\mathbf{P}^0, \mathbf{Q}^0\}$ denote embedding matrices;

\WHILE{not converged}
    \STATE Sample a batch training data $\mathcal{D}$;
    \STATE \textit{\textbf{\# Environment Simulation \#}}
    \STATE Compute the edge drop probability~(Eq.~\eqref{eq: drop probability});
    \STATE Obtain edge drop distribution by Gumbel-Softmax~( Eq.\eqref{eq: drop reparameter});
    \STATE Extract $K$ social subgraphs~(Eq.~\eqref{eq: subgraph extraction});
    \STATE \textit{\textbf{\# Invariant Preference Learning \#}}
    \STATE Learn user preference in each environment~(Eq.\eqref{eq:adj_matrix}-Eq.\eqref{eq:readout});
    \STATE Obtain ERM loss in each environment~(Eq.\eqref{eq: SSL});
    \STATE Obtain overall loss across all environments~(Eq.\eqref{eq: overall loss});
    \STATE \textit{\textbf{\# Adversarial Training \#}}
    \STATE Update model parameters $\theta, \phi$ according to invariance based loss minimization~(Eq.\eqref{eq: optimization});
    \STATE Update environment generator parameters $\theta$ according to invariance loss maximization~(Eq.(17))) 
\ENDWHILE

\RETURN the optimal $\mathcal{G}^{*}_{\theta, \phi}(\cdot)$
\end{algorithmic}
\label{alg: SGIL}
\end{algorithm}

\subsection{Model Discussion}
In this section, we analyze the proposed \shortname~from both space and time complexity.

\subsubsection{Space Complexity}
As illustrated in Algorithm 1, the parameters of \shortname~consist of two components: the embedding parameters $\phi$ and the environment generator parameters $\theta$. The embedding parameters $\phi \in \mathbb{R}^{(M+N) \times d}$ are the same as those used in general social recommenders~(such as DiffNet\cite{DiffNet} and GBSR~\cite{yang2024graph}). In addition, the environment generator parameters $\theta$ consist of $K$ individual MLPs~($\mathbb{R}^{2d \times d} + \mathbb{R}^{d \times 1}$). Since the embedding size $d$ is much smaller than node number $M+N$, the additional parameters introduced by \shortname~is negligible.

\subsubsection{Time Complexity}
The computational overhead introduced by \shortname~mainly consists of three components: environment simulation, invariant preference learning, and adversarial training. For environment simulation, we first obtain the encoded node representations using GCNs, and then employ $K$ individual MLPs to generate diverse environments. For preference learning, we learn user and item representations within each environment and optimize them through an invariance-based objective. Finally, we adopt an across-batch adversarial training strategy to iteratively update the parameters. Compared to the backbone model (LightGCN-S), \shortname~incurs approximately $K$ times the computational cost during the preference learning stage, and the employed softmax-base loss\cite{wu2024effectiveness} is computationally more intensive than the traditional BPR loss~\cite{UAI2009BPR}. Fortunately, in our experiments, the optimal value of $K$ is typically less than 5, and \shortname~converges much faster than LightGCN-S. Overall, the total training time of \shortname~is approximately 1 to 3 times that of LightGCN-S.

\begin{table}[th]
    \centering
	\setlength{\belowcaptionskip}{5pt} %
	\caption{The statistics of three datasets.}\label{tab:statistics}
	\vspace{-0.2cm}
	\scalebox{0.95}{
    \begin{tabular}{c|c|c|c}
    \hline
    Dataset & Douban-Book & Yelp & Epinions \\ \hline
    Users & 13,024 & 19,593 & 18,202  \\ 
    Items & 22,347 & 21,266 & 47,449  \\ \hline
    Interactions & 792,062 & 450,884 & 298,173 \\ 
    Social Relations & 169,150 & 864,157 & 381,559 \\ \hline
    Interaction Density & 0.048\%  & 0.034\%  & 0.035\%  \\
    Relation Density & 0.268\% & 0.206\% & 0.115\% \\ \hline
    \end{tabular}}
    \vspace{-0.2cm}
\end{table}

\section{Experiments}
\subsection{Experimental Settings}
\subsubsection{Datasets} Following~\cite{yang2024graph}, we select three public social recommendation datasets to conduct our empirical studies: Douban-Book, Yelp, and Epinions. Each dataset includes user-user social links and user-item interactions. For our experiments, we randomly sample 80\% of the interactions as training data, with the remaining 20\% reserved as test data. The detailed statistics of all datasets are summarized in Table \ref{tab:statistics}.

\subsubsection{Baselines and Evaluation Metrics.}
We compare \textit{SGIL} with state-of-the-art methods. The baselines can be divided into two groups: 1) Graph-based social recommendations: GraphRec~\cite{GraphRec}, DiffNet++~\cite{wu2020diffnet++}, SocialLGN~\cite{liao2022sociallgn}, LightGCN-S~\cite{LightGCN}; 2) Social graph denoising methods: ESRF~\cite{TKDE2020enhancing}, GDMSR~\cite{WWW2023robust}, SEPT~\cite{yu2021self},  ShaRe~\cite{jiang2024challenging}, GBSR~\cite{yang2024graph}, which are list as follows:
\begin{itemize}[leftmargin=1em]
    \item \textbf{LightGCN}~\cite{LightGCN}: is the SOTA graph-based collaborative filtering method, which simplifies GCNs by removing the redundant feature transformation and non-linear activation components for ID-based recommendation.
    \item \textbf{LightGCN-S}~\cite{LightGCN, yang2024graph}: extends LightGCN to graph-based social recommendation, that each user's neighbors include their interacted items and linked social users.
    \item \textbf{GraphRec}~\cite{GraphRec}: is a classic graph-based social recommendation method, it incorporates user opinions and user two kinds of graphs for preference learning.
    \item \textbf{DiffNet++}~\cite{wu2020diffnet++}: is the SOTA graph-based social recommendation method, it recursively formulates user interest propagation and social influence diffusion process with a hierarchical attention mechanism.
    \item \textbf{SocialLGN}~\cite{liao2022sociallgn}: propagates user representations on both user-item interactions graph and user-user social graph with light graph convolutional layers, and fuses them for recommendation.
    \item \textbf{ESRF}~\cite{TKDE2020enhancing}: generates alternative social neighbors and further performs neighbor denoising with adversarial training.
    \item \textbf{GDMSR}~\cite{WWW2023robust}: designs the robust preference-guided social denoising to enhance graph-based social recommendation, it only remains the informative social relations according to preference confidences.
    \item \textbf{SEPT}~\cite{yu2021self}: is the SOTA self-supervised social recommendation method, it incorporates social relations to contrastive learning.
    \item \textbf{SHaRe}~\cite{jiang2024challenging}: challenges the low social homophily and proposes a social graph rewriting strategy with relation-based contrastive learning enhancement.
    \item \textbf{GBSR}~\cite{yang2024graph}: devises a model-agnostic social denoising method, which leverages the information bottleneck principle to remove the redundant social structure for recommendation tasks.
\end{itemize}

The evaluation metrics are Recall@N and NDCG@N~\cite{jarvelin2002cumulated}, which have been widely used in ranking-based recommender systems. Following the mainstream evaluation process~\cite{zhao2020revisiting}, we adopt a full-ranking protocol that all non-interacted items are candidates for evaluation.

\subsubsection{Implementation Details} 
All embedding-based methods are initialized by a normal initializer with a mean value of 0 and a standard variance of 0.01, the embedding size is fixed to 64. For model optimization, we use the Adam optimizer~\cite{kingma2014adam} with a learning rate of 0.001 and a batch size of 2048. We empirically set the GCN layer $L=3$ and the temperature coefficient $\tau=0.2$ for all datasets. For the edge observation bias, we follow the previous work~\cite{yang2024graph} and set the bias $\epsilon=0.5$ for all datasets. For the environment number $K$, the variance-based penalty coefficient $\beta$, we carefully search the best parameters for each dataset and report the detailed comparisons. For the adversarial training step, we set $T=20$ on both the Douban-Book and Yelp datasets, and $T=3$ on the Epinions datasets. For all baselines, we refer to the original parameters and carefully fine-tune them for fair comparisons.

\begin{table*}[t]
\centering
\caption{Overall performance comparisons on three benchmarks~($\textit{p-value}<0.05$). The best performance is highlighted in \textbf{bold} and the second is highlighted by \underline{underlines}.} 
 \vspace{-2pt}
\label{tab: overall performance}
\scalebox{1.0}{
\begin{tabular}{|l|c|c|c|c|c|c|c|c|c|c|c|c|}
\hline 
 & \multicolumn{4}{c|}{Douban-Book}       
 & \multicolumn{4}{c|}{Yelp}
 & \multicolumn{4}{c|}{Epinions}\\ \cline{2-13}
\multirow{-2}{*}{Models} &R@10&N@10&R@20&N@20 &R@10&N@10&R@20&N@20 &R@10&N@10&R@20&N@20  \\ \hline
LightGCN  & 0.1039 & 0.1195 & 0.1526 & 0.1283 & 0.0698 & 0.0507 & 0.1081 & 0.0623 & 0.0432 & 0.0314 & 0.0675 & 0.0385 \\ \hline
GraphRec & 0.0971 & 0.1145 & 0.1453 & 0.1237 & 0.0672 & 0.0485 & 0.1077 & 0.0607 & 0.0436 & 0.0315 & 0.0681 & 0.0387 \\ \hline
DiffNet++ & 0.1010 & 0.1184 & 0.1489 & 0.1270 & 0.0707 & 0.0516 & 0.1114 & 0.0640 & 0.0468 & 0.0329 & 0.0727 & 0.0406 \\ \hline 
SocialLGN & 0.1034 & 0.1182 & 0.1527 & 0.1274 & 0.0681 & 0.0507 & 0.1059 & 0.0620 & 0.0416 & 0.0307 & 0.0634 & 0.0371 \\ \hline
LightGCN-S & 0.1021 & 0.1187 & 0.1506 & 0.1281 & 0.0714 & 0.0529 & 0.1126 & 0.0651 & 0.0477 & 0.0347 & 0.0716 & 0.0417  \\ \hline \hline
ESRF & 0.1042 & 0.1199 & 0.1534 & 0.1301 & 0.0718 & 0.0526 & 0.1123 & 0.0645 & 0.0462 & 0.0329 & 0.0727 & 0.0406 \\ \hline
GDMSR  & 0.1026 & 0.1001 & 0.1538 & 0.1245 & 0.0739 & 0.0535 & 0.1148 & 0.0658 & 0.0461 & 0.0326 & 0.0721 & 0.0414 \\ \hline
SEPT & 0.1094 & 0.1300 & 0.1592 & 0.1382 & 0.0749 & 0.0553 & 0.1176 & 0.0682 & 0.0457 & 0.0341 & 0.0724 & 0.0416 \\ \hline
SHaRe & 0.1050 & 0.1243 & 0.1544 & 01325 & 0.0724 & 0.0534 & 0.1138 & 0.0658 & 0.0490 & 0.0354 & 0.0771 & 0.0438\\ \hline
GBSR  & \underline{0.1189} & \underline{0.1451} & \underline{0.1694} & \underline{0.1523} & \underline{0.0805} & \underline{0.0592} & \underline{0.1243} & \underline{0.0724} & \underline{0.0529} & \underline{0.0385} & \underline{0.0793} & \underline{0.0464} \\ \hline
SGIL & \textbf{0.1303} & \textbf{0.1567} & \textbf{0.1809} & \textbf{0.1627} & \textbf{0.0872} & \textbf{0.0654} & \textbf{0.1323} & \textbf{0.0787} & \textbf{0.0565} & \textbf{0.0409} & \textbf{0.0844} & \textbf{0.0489} \\ \hline
\textbf{Impro.} & \textbf{9.59\%} & \textbf{7.99\%} & \textbf{6.79\%} & \textbf{6.83\%} & \textbf{8.93\%} & \textbf{8.32\%} & \textbf{6.44\%} & \textbf{8.70\%} & \textbf{6.81\%} & \textbf{6.23\%} & \textbf{6.43\%} & \textbf{5.39\%}  \\ \hline
\end{tabular}}
\vspace{-0.2cm}
\end{table*}

\subsection{Performance Comparisons}
Following the current SOTA method~\cite{yang2024graph}, we implement all social denoising methods on the LightGCN-S backbone. As shown in Table 2, we have the following observations:

\begin{itemize}[leftmargin=*] 
\item \textbf{Compared to LightGCN, graph-based social recommendation methods show modest improvements on most datasets.} For instance, DiffNet++ achieves a 2.24\% gain in NDCG@20 on the Yelp dataset. However, this is not universally observed; all graph-based social recommendation methods experience performance degradation on the Douban-Book dataset. These results highlight an important observation: while social graphs provide additional information, graph-based social recommendation methods do not consistently outperform LightGCN in terms of performance. This suggests that directly incorporating social graphs may sometimes hinder recommendation performance, emphasizing the need to eliminate redundant social relations to improve outcomes.
\item \textbf{Social denoising methods outperform traditional graph-based methods.} Compared to the backbone model (LightGCN-S), we observe that social denoising methods achieve stable improvements across all three datasets. Whether leveraging adversarial training (ESRF), preference-guided graph filtering (GDMSR), self-supervised signals (SEPT and SHaRe), or the graph information bottleneck principle (GBSR), these methods demonstrate remarkable performance enhancements. Notably, GBSR stands out as the strongest baseline among them, effectively preserving the minimal yet sufficient social structure necessary for recommendation tasks. By eliminating redundant information while retaining valuable social connections, GBSR effectively mitigates the impact of social noise, thereby enhancing recommendation accuracy. In summary, learning a denoised social graph significantly boosts the effectiveness of social recommendation, underscoring the critical role of social denoising in practical applications.
\item \textbf{Our proposed \shortname~consistently outperforms all baselines by a substantial margin.} Specifically, \shortname~improves upon the strongest baseline (GBSR) to NDCG@10 by 7.99\%, 10.47\%, and 5.97\% gains on the Douban-Book, Yelp, and Epinions datasets, respectively. These experimental results strongly validate the superiority of our proposed \shortname~model. Compared to the backbone model (LightGCN-S), \shortname\ achieves significant improvements of approximately 32.01\%, 23.63\%, and 17.58\% in terms of NDCG@20 across three benchmark datasets. Experimental results verify the effectiveness of our proposed \shortname~. Compared to traditional graph-based social recommendation methods,
\shortname~effectively alleviates the adverse effects of social noise and enhances recommendation performance. Compared to current social denoising methods, \shortname~simulates diverse social noise environments and learns invariant user preferences, providing stronger guarantees under various potential noise scenarios. This innovative approach not only improves the robustness of the recommendation system but also offers a novel solution for complex social recommendation contexts, ensuring stability and reliability when faced with social noise.
\end{itemize}

\begin{table*}[t]
\centering
\caption{Ablation study of \shortname~on three datasets.}
 \vspace{-5pt}
\label{tab: ablation study}
\scalebox{0.98}{
\begin{tabular}{l|cccc|cccc|cccc}
\hline 
 & \multicolumn{4}{c}{Douban-Book}       
 & \multicolumn{4}{c}{Yelp}
 & \multicolumn{4}{c}{Epinions}\\ \cline{2-13}
\multirow{-2}{*}{Models} &R@10&N@10&R@20&N@20 &R@10&N@10&R@20&N@20 &R@10&N@10&R@20&N@20  \\ \hline
\textit{SGIL}-w/o EG & 0.1243 & 0.1445 & 0.1672 & 0.1495 & 0.0795 & 0.0591 & 0.1201 & 0.0713 & 0.0475 & 0.0348 & 0.0708 & 0.0417 \\ \hline
\textit{SGIL}-w/o IL & 0.1272 & 0.1503 & 0.1748 & 0.1562 & 0.0813 & 0.0612 & 0.1229 & 0.0735 & 0.0499 & 0.0361 & 0.0749 & 0.0436 \\ \hline
\textit{SGIL}-w/o EE & 0.1292 & 0.1562 & 0.1804 & 0.1626 & 0.0850 & 0.0643 & 0.1290 & 0.0773 & 0.0556 & 0.0410 & 0.0825 & 0.0487 \\ \hline
\textbf{\textit{SGIL}} & \textbf{0.1303} & \textbf{0.1567} & \textbf{0.1809} & \textbf{0.1627} & \textbf{0.0872} & \textbf{0.0654} & \textbf{0.1323} & \textbf{0.0787} & \textbf{0.0565} & \textbf{0.0409} & \textbf{0.0844} & \textbf{0.0489} \\ \hline
\end{tabular}}
\end{table*}

\subsection{Ablation Study}
To further analyze the effect of each component of the proposed \shortname~, we conduct ablation studies on three datasets. As shown in Table~\ref{tab: ablation study}, we compare \shortname~with three corresponding variants: 
\textbf{\textit{SGIL}-w/o EG} denotes that without the environment generator, then environment-inspired invariant learning disappears and \shortname~degenerates to the backbone model~(LightGCN-S+ERM).
\textbf{\textit{SGIL}-w/o IL} denotes that without the invariant learning regularizer, we only adopt the mean representations of all environments for inference. 
\textbf{\textit{SGIL}-w/o EE} denotes that without environment exploration, we only perform graph invariant learning on the initialized environments.
From Table~\ref{tab: ablation study}, we have the following observations. First, \textit{SGIL}-w/o EG exhibits the most performance degeneration, which indicates the importance of environment simulation. Environment simulation is the basis of \shortname~, as the task-irrelevant relations are unknown, so \shortname~simulates diverse environments and encourages preference consistency across environments to alleviate the effect of potential noise.
Second, \textit{SGIL}-w/o IL also performs significant degradation, which demonstrates the effectiveness of the invariance-based regularizer. Finally, \textit{SGIL}-w/o EE shows a consistent performance drop. Because social noise is not directly perceived, we use an adversarial training strategy to encourage more diverse environments. The more diverse the environments the better the robustness of the proposed method. All ablation results demonstrate the effectiveness of each technical component of our proposed \shortname~.

\begin{figure} [t]
    \centering
    \subfloat[Douban-Book]{
    \includegraphics[width=42mm]{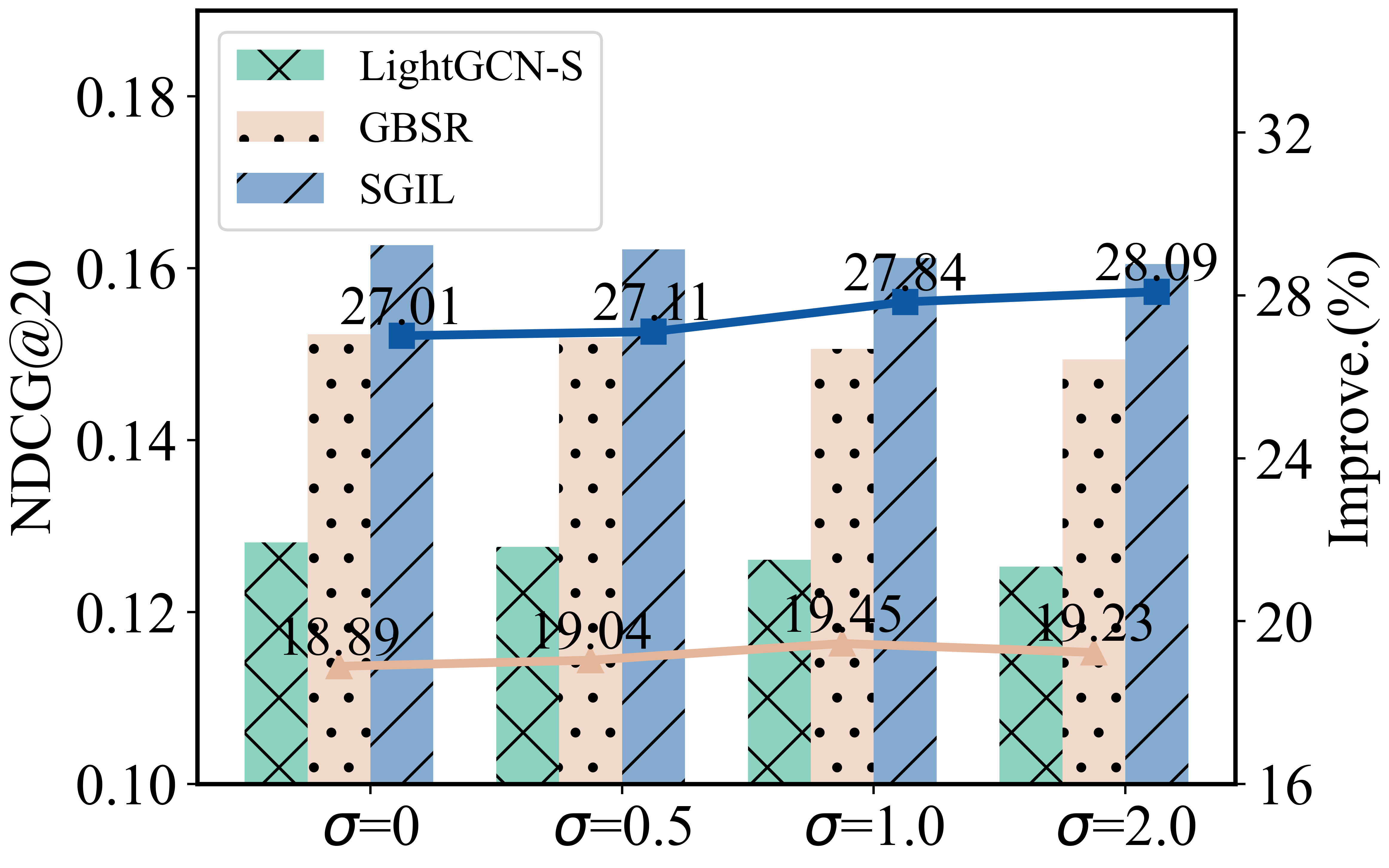}}
    \subfloat[Yelp]{\includegraphics[width=42mm]{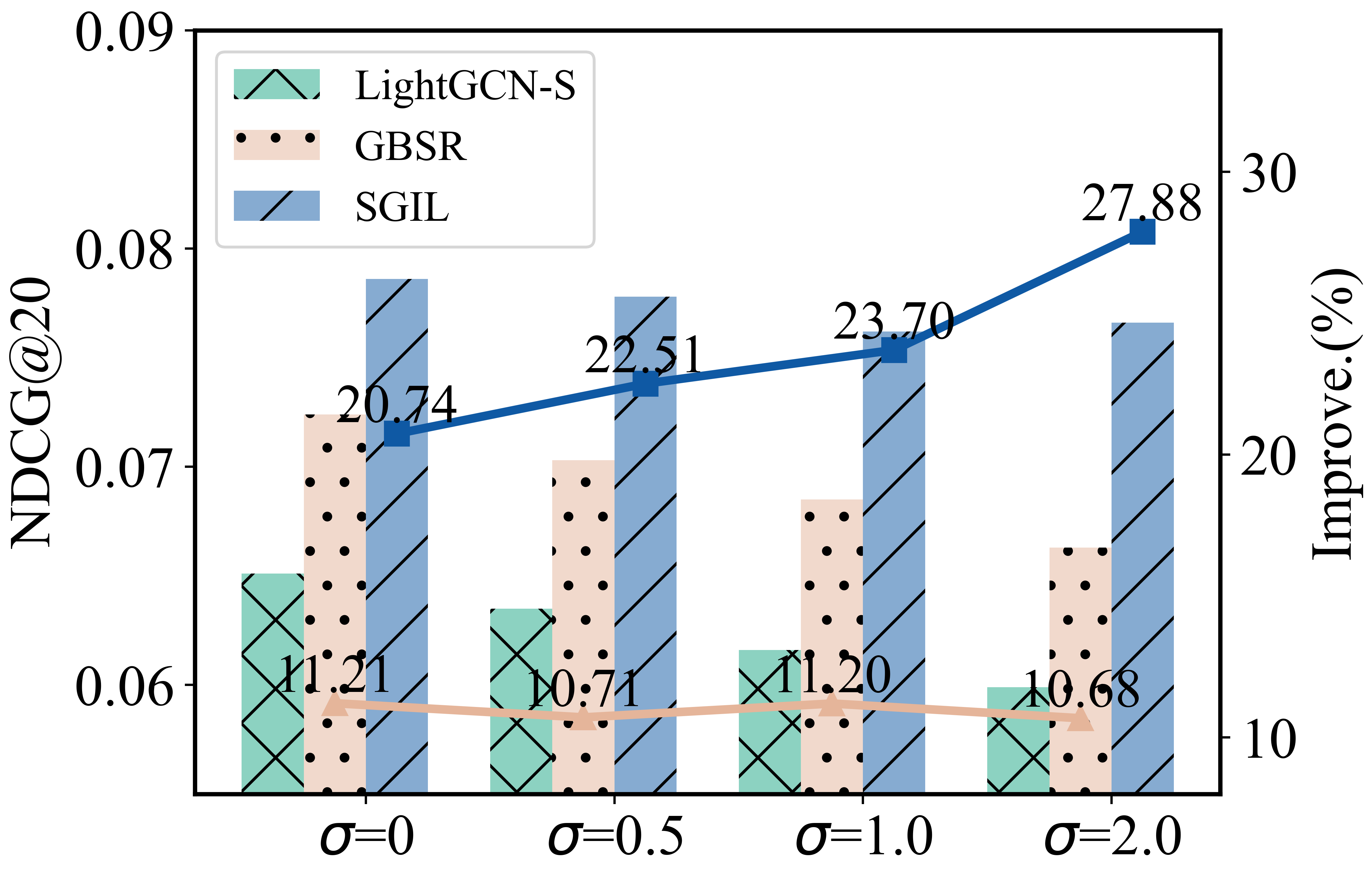}}
    \vspace{-0.2cm}
    \caption{Recommendation performance~(NDCG@20) under different degrees of noise.}
    \label{fig: noise scale}
    \vspace{-0.4cm}
\end{figure}

\subsection{Detailed Analysis of \shortname~}
In this section, we further analyze \shortname~from the following aspects: robustness to different scale noises, sparsity analysis, and parameter sensitivities. 
\subsubsection{A-Robustness to different noises.} 
To evaluate the robustness under different degrees of noise, we conduct experiments on the semi-synthetic datasets, in which we inject different proportions of fake social relations into the original social graph. As illustrated in Figure \ref{fig: noise scale}, we exhibit the recommendation performances under different degrees of noise, where the horizontal axis $\sigma$ represents the injection fake relation ratio, the vertical axes denote NDCG@20 and relative improvements compared with the backbone model.
We can find that with the increase in noise degree, all models show a performance decrease. However, social denoising methods perform higher improvements than the backbone model when the noise degree increases. This verifies that social denoising methods have better robustness for highly noisy environments. Furthermore, our proposed \shortname~consistently shows significant improvements with various settings, especially improving the backbone model about $28\%$ gains when the noise ratio $\sigma=2.0$. Empowering the ability of graph invariant learning, \shortname~can extract invariant social structure for robust preference learning.

\begin{figure} [t]
    \centering
    \subfloat[Douban-Book]{
    \includegraphics[width=40mm]{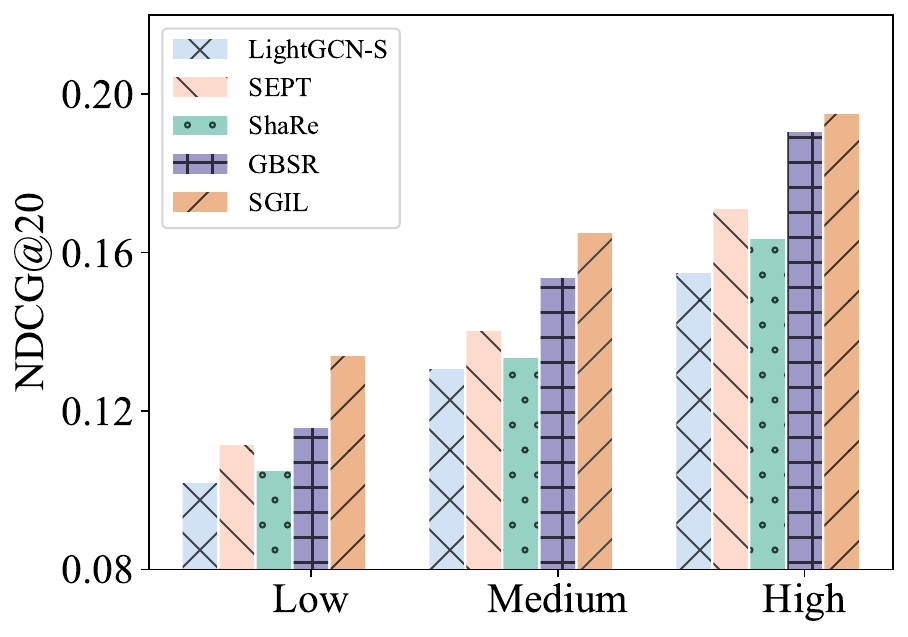}}
    \subfloat[Yelp]{\includegraphics[width=40mm]{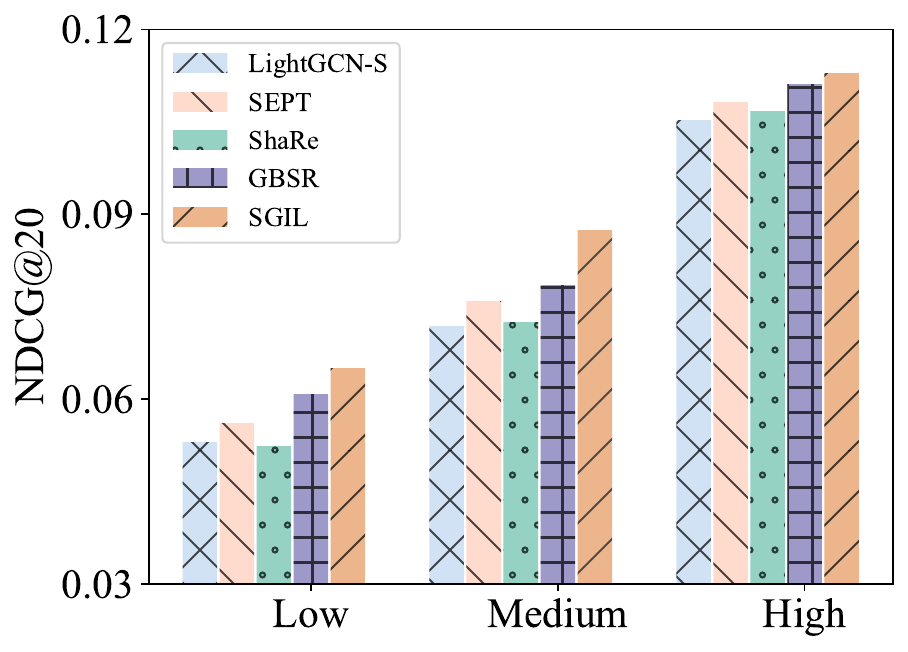}}
    \vspace{-0.2cm}
    \caption{Recommendation performance~(NDCG@20) across different sparsity users.}
    \vspace{-0.3cm}
    \label{fig: sparsity analysis}
\end{figure}

\subsubsection{B-Sparsity Analysis.}
Here, we investigate how the proposed \shortname~performs for different sparsity users. As depicted in Figure \ref{fig: sparsity analysis}, we report recommendation performances of \shortname~and other social denoising methods on different sparsity user groups. Specifically, we divide all users into 'Low, Medium, High' three groups based on their historical interaction records. Then, we evaluate all methods across different groups. From Figure \ref{fig: sparsity analysis}, we observe that with the increase in interaction density, each model shows significant performance improvements. Compared with the backbone model, our proposed \shortname~consistently presents better performances in all settings. Especially, \shortname~performs higher improvements on the sparser scenarios, because their long-tail users are more dependent on social relations, our proposed \shortname~provides more stable user preferences in the noisy environment.

\subsubsection{C-Parameter Sensitivities.}
\begin{figure} [t]
    \centering
    \subfloat{\includegraphics[width=28mm]{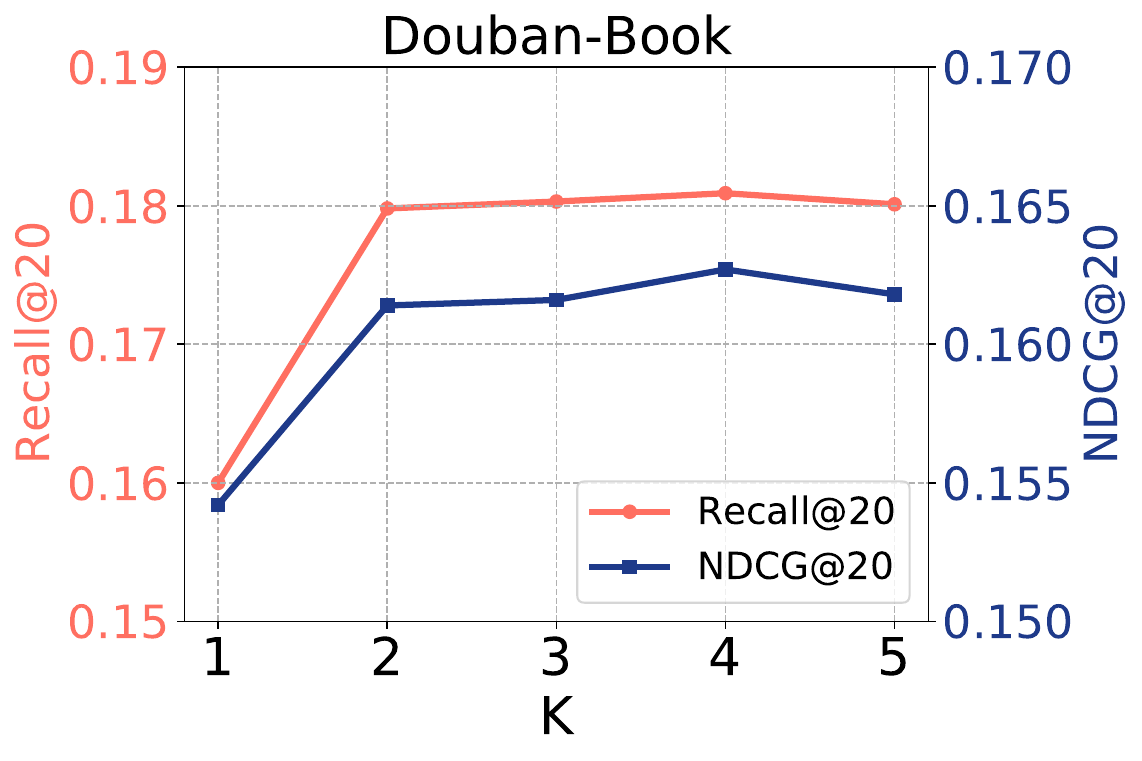}}
    \subfloat{\includegraphics[width=28mm]{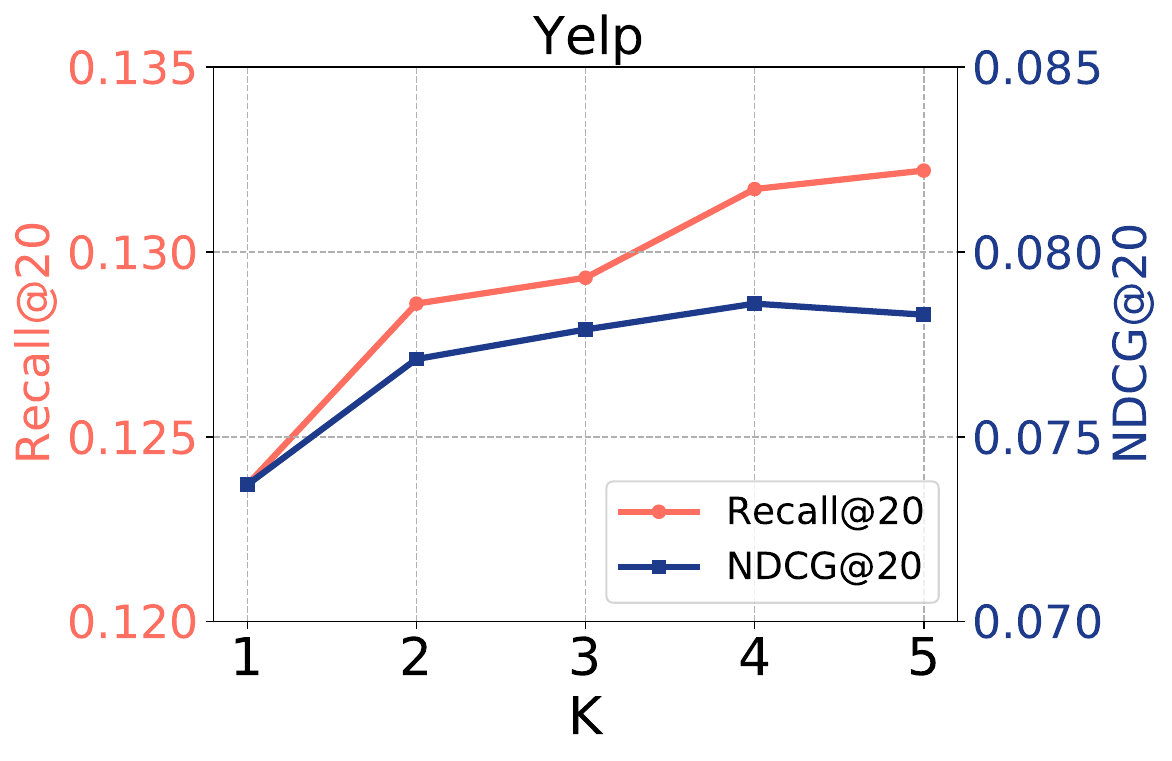}}
    \subfloat{\includegraphics[width=28mm]{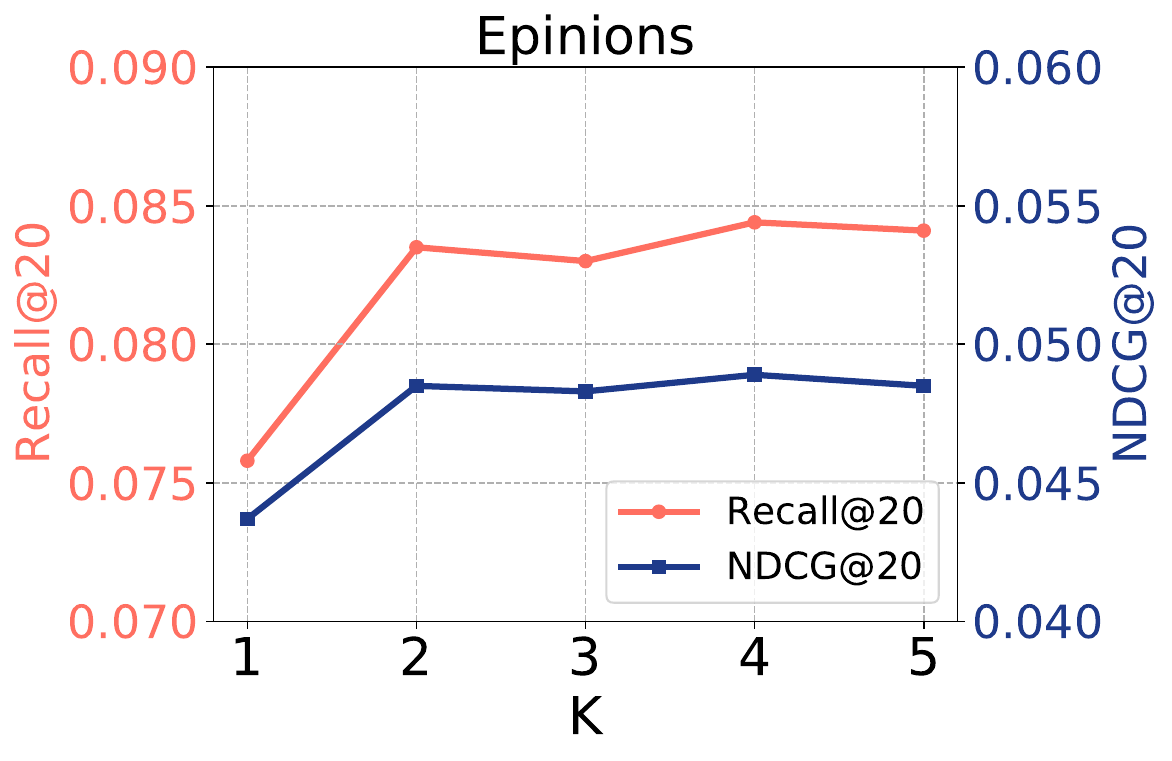}} \\
    \vspace{-0.3cm}
    \subfloat{\includegraphics[width=28mm]{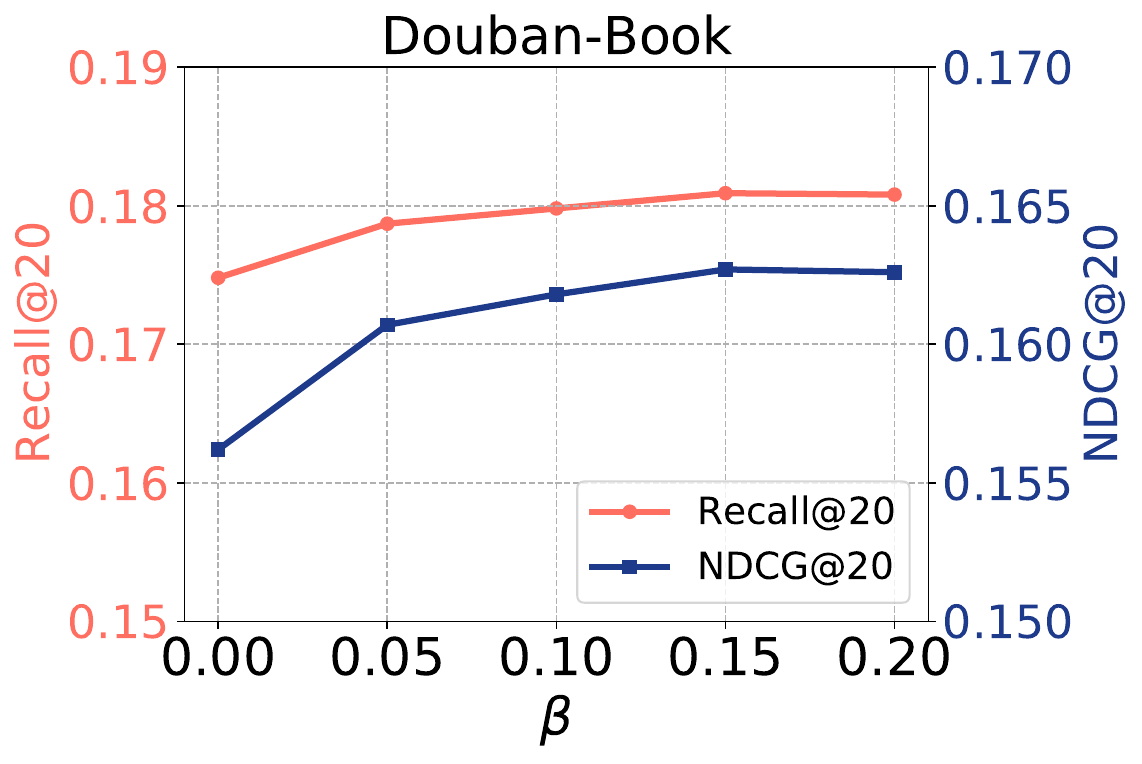}}
    \subfloat{\includegraphics[width=28mm]{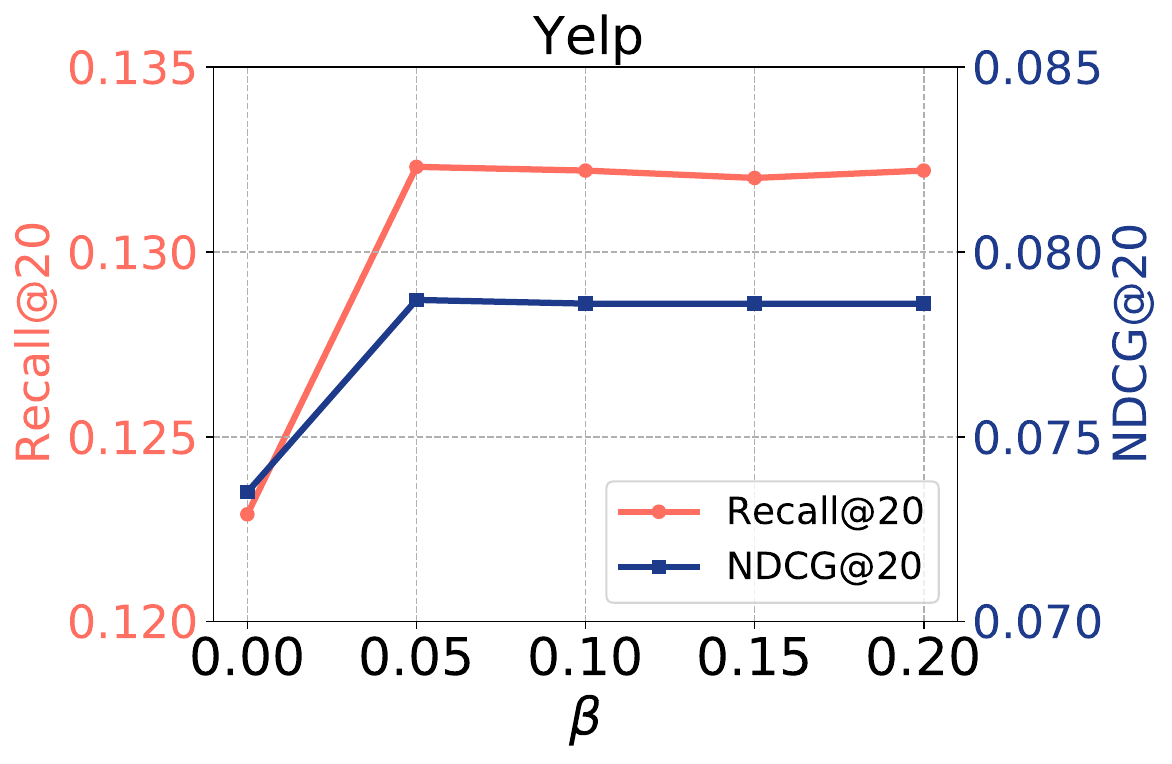}}
    \subfloat{\includegraphics[width=28mm]{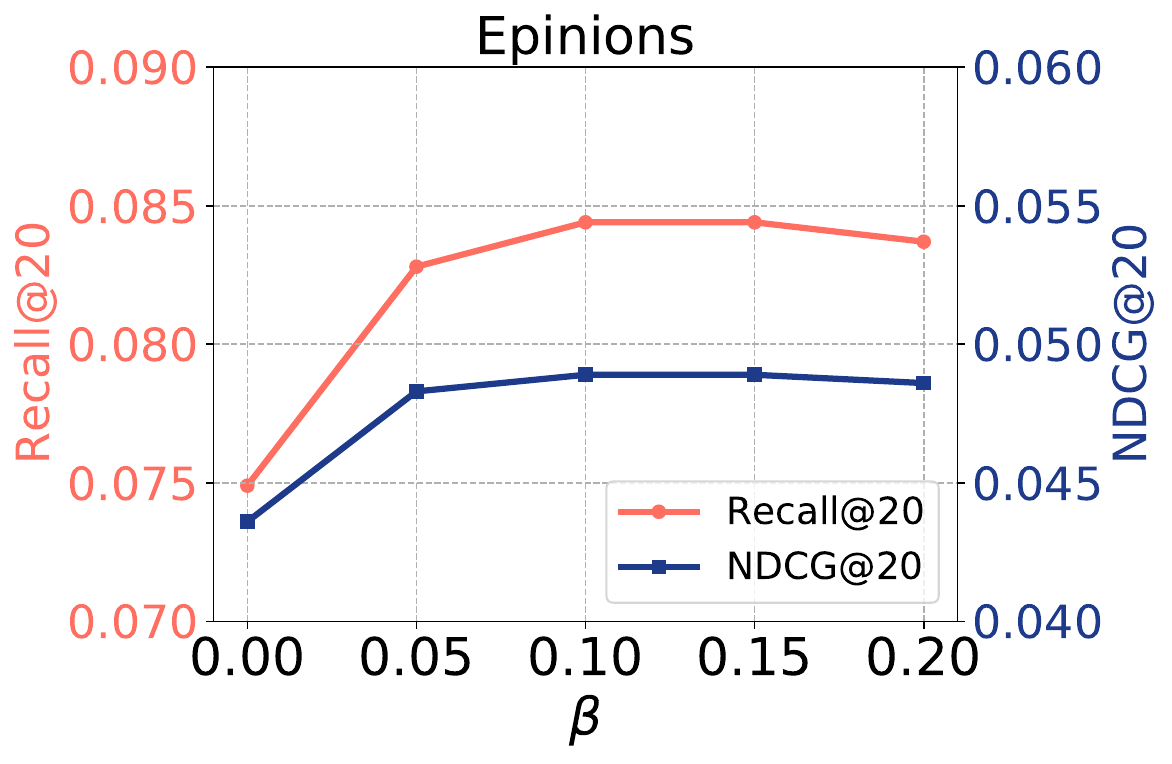}}
    \vspace{-0.3cm}
    \caption{Impact of the number of environments $K$ and invariance coefficient $\beta$.}
    \label{fig: parmaeters}
    \vspace{-0.2cm}
\end{figure}

We next analyze the influence of two core parameters in \shortname~: the environment number $K$ and the invariance penalty coefficient $\beta$. We search these two parameters carefully and report the performances with different parameters on three datasets. The parameter $K$ controls the exploration range and the coefficient $\beta$ determines the weight of the invariance constraint. As shown in Figure \ref{fig: parmaeters}, we search $K \in \{1,2,3,4,5\}$ and $\beta \in \{0, 0.05, 0.10, 0.15, 0.20\}$. In which, $K=1$ denotes that we only perform a single environment generator, then the invariance constraint disappears. Besides, $\beta=0$ also denotes the invariance constraint disappears, the difference compared with $K=0$ is the number of environments is not the same. We can observe that \shortname~achieves the optimal results when $(K=4, \beta=0.15)$ on the Douban-Book dataset, $(K=4, \beta=0.05)$ on the Yelp dataset, and $(K=4, \beta=0.10)$ on the Epinions dataset.

\section{Conclusion}
In this paper, we investigate social denoising recommendations from a novel invariant learning perspective. Technically, we propose a novel \fullname~approach, aiming to learn the invariant user preference across diverse noisy environments. To achieve this goal, we first simulate multiple social environments based on user preferences guidance, and then learn invariant user preferences from invariance-based optimization. Particularly, we derive the empirical risk minimization objective in each environment when considering the characterization of recommendation tasks. 
Furthermore, \shortname~fully explores diverse environments with an adversarial training strategy. This strategy allows the model to adapt to a wide range of challenging conditions, thereby improving its generalization and robustness. Extensive experiments on three benchmarks verify the effectiveness of the proposed \shortname~. 

\begin{acks}
This work was supported in part by grants from the National Key Research and Development Program of China (Grant No. 2021ZD0111802), the National Natural Science Foundation of China (Grant No. U23B2031, 721881011).
\end{acks}

\balance
\bibliographystyle{ACM-Reference-Format}
\bibliography{SGIL} 


\begin{thebibliography}{50}


\ifx \showCODEN    \undefined \def \showCODEN     #1{\unskip}     \fi
\ifx \showDOI      \undefined \def \showDOI       #1{#1}\fi
\ifx \showISBNx    \undefined \def \showISBNx     #1{\unskip}     \fi
\ifx \showISBNxiii \undefined \def \showISBNxiii  #1{\unskip}     \fi
\ifx \showISSN     \undefined \def \showISSN      #1{\unskip}     \fi
\ifx \showLCCN     \undefined \def \showLCCN      #1{\unskip}     \fi
\ifx \shownote     \undefined \def \shownote      #1{#1}          \fi
\ifx \showarticletitle \undefined \def \showarticletitle #1{#1}   \fi
\ifx \showURL      \undefined \def \showURL       {\relax}        \fi
\providecommand\bibfield[2]{#2}
\providecommand\bibinfo[2]{#2}
\providecommand\natexlab[1]{#1}
\providecommand\showeprint[2][]{arXiv:#2}

\bibitem[Arjovsky et~al\mbox{.}(2019)]%
        {IRM}
\bibfield{author}{\bibinfo{person}{Martin Arjovsky}, \bibinfo{person}{L{\'e}on Bottou}, \bibinfo{person}{Ishaan Gulrajani}, {and} \bibinfo{person}{David Lopez-Paz}.} \bibinfo{year}{2019}\natexlab{}.
\newblock \showarticletitle{Invariant risk minimization}.
\newblock \bibinfo{journal}{\emph{arXiv preprint arXiv:1907.02893}} (\bibinfo{year}{2019}).
\newblock


\bibitem[Bai et~al\mbox{.}(2024)]%
        {SIGIR2024multimodality}
\bibfield{author}{\bibinfo{person}{Haoyue Bai}, \bibinfo{person}{Le Wu}, \bibinfo{person}{Min Hou}, \bibinfo{person}{Miaomiao Cai}, \bibinfo{person}{Zhuangzhuang He}, \bibinfo{person}{Yuyang Zhou}, \bibinfo{person}{Richang Hong}, {and} \bibinfo{person}{Meng Wang}.} \bibinfo{year}{2024}\natexlab{}.
\newblock \showarticletitle{Multimodality Invariant Learning for Multimedia-Based New Item Recommendation}. In \bibinfo{booktitle}{\emph{SIGIR}}. \bibinfo{pages}{677--686}.
\newblock


\bibitem[Cao(2016)]%
        {cao2016non}
\bibfield{author}{\bibinfo{person}{Longbing Cao}.} \bibinfo{year}{2016}\natexlab{}.
\newblock \showarticletitle{Non-iid recommender systems: A review and framework of recommendation paradigm shifting}.
\newblock \bibinfo{journal}{\emph{Engineering}} \bibinfo{volume}{2}, \bibinfo{number}{2} (\bibinfo{year}{2016}), \bibinfo{pages}{212--224}.
\newblock


\bibitem[Chen et~al\mbox{.}(2022)]%
        {chen2022learning}
\bibfield{author}{\bibinfo{person}{Yongqiang Chen}, \bibinfo{person}{Yonggang Zhang}, \bibinfo{person}{Yatao Bian}, \bibinfo{person}{Han Yang}, \bibinfo{person}{MA Kaili}, \bibinfo{person}{Binghui Xie}, \bibinfo{person}{Tongliang Liu}, \bibinfo{person}{Bo Han}, {and} \bibinfo{person}{James Cheng}.} \bibinfo{year}{2022}\natexlab{}.
\newblock \showarticletitle{Learning causally invariant representations for out-of-distribution generalization on graphs}.
\newblock \bibinfo{journal}{\emph{NeurIPS}}  \bibinfo{volume}{35} (\bibinfo{year}{2022}), \bibinfo{pages}{22131--22148}.
\newblock


\bibitem[Creager et~al\mbox{.}(2021)]%
        {EIIL}
\bibfield{author}{\bibinfo{person}{Elliot Creager}, \bibinfo{person}{J{\"o}rn-Henrik Jacobsen}, {and} \bibinfo{person}{Richard Zemel}.} \bibinfo{year}{2021}\natexlab{}.
\newblock \showarticletitle{Environment inference for invariant learning}. In \bibinfo{booktitle}{\emph{ICML}}. PMLR, \bibinfo{pages}{2189--2200}.
\newblock


\bibitem[Du et~al\mbox{.}(2022)]%
        {MM2022invariant}
\bibfield{author}{\bibinfo{person}{Xiaoyu Du}, \bibinfo{person}{Zike Wu}, \bibinfo{person}{Fuli Feng}, \bibinfo{person}{Xiangnan He}, {and} \bibinfo{person}{Jinhui Tang}.} \bibinfo{year}{2022}\natexlab{}.
\newblock \showarticletitle{Invariant representation learning for multimedia recommendation}. In \bibinfo{booktitle}{\emph{MM}}. \bibinfo{pages}{619--628}.
\newblock


\bibitem[Fan et~al\mbox{.}(2019)]%
        {GraphRec}
\bibfield{author}{\bibinfo{person}{Wenqi Fan}, \bibinfo{person}{Yao Ma}, \bibinfo{person}{Qing Li}, \bibinfo{person}{Yuan He}, \bibinfo{person}{Eric Zhao}, \bibinfo{person}{Jiliang Tang}, {and} \bibinfo{person}{Dawei Yin}.} \bibinfo{year}{2019}\natexlab{}.
\newblock \showarticletitle{Graph neural networks for social recommendation}. In \bibinfo{booktitle}{\emph{TheWebConf}}. \bibinfo{pages}{417--426}.
\newblock


\bibitem[Guo et~al\mbox{.}(2015)]%
        {guo2015trustsvd}
\bibfield{author}{\bibinfo{person}{Guibing Guo}, \bibinfo{person}{Jie Zhang}, {and} \bibinfo{person}{Neil Yorke-Smith}.} \bibinfo{year}{2015}\natexlab{}.
\newblock \showarticletitle{Trustsvd: Collaborative filtering with both the explicit and implicit influence of user trust and of item ratings}. In \bibinfo{booktitle}{\emph{AAAI}}, Vol.~\bibinfo{volume}{29}.
\newblock


\bibitem[He et~al\mbox{.}(2020)]%
        {LightGCN}
\bibfield{author}{\bibinfo{person}{Xiangnan He}, \bibinfo{person}{Kuan Deng}, \bibinfo{person}{Xiang Wang}, \bibinfo{person}{Yan Li}, \bibinfo{person}{Yongdong Zhang}, {and} \bibinfo{person}{Meng Wang}.} \bibinfo{year}{2020}\natexlab{}.
\newblock \showarticletitle{Lightgcn: Simplifying and powering graph convolution network for recommendation}. In \bibinfo{booktitle}{\emph{SIGIR}}. \bibinfo{pages}{639--648}.
\newblock


\bibitem[He et~al\mbox{.}(2024)]%
        {he2024double}
\bibfield{author}{\bibinfo{person}{Zhuangzhuang He}, \bibinfo{person}{Yifan Wang}, \bibinfo{person}{Yonghui Yang}, \bibinfo{person}{Peijie Sun}, \bibinfo{person}{Le Wu}, \bibinfo{person}{Haoyue Bai}, \bibinfo{person}{Jinqi Gong}, \bibinfo{person}{Richang Hong}, {and} \bibinfo{person}{Min Zhang}.} \bibinfo{year}{2024}\natexlab{}.
\newblock \showarticletitle{Double Correction Framework for Denoising Recommendation}.
\newblock \bibinfo{journal}{\emph{KDD}} (\bibinfo{year}{2024}).
\newblock


\bibitem[Hu et~al\mbox{.}(2025)]%
        {hu2025modality}
\bibfield{author}{\bibinfo{person}{Jun Hu}, \bibinfo{person}{Bryan Hooi}, \bibinfo{person}{Bingsheng He}, {and} \bibinfo{person}{Yinwei Wei}.} \bibinfo{year}{2025}\natexlab{}.
\newblock \showarticletitle{Modality-Independent Graph Neural Networks with Global Transformers for Multimodal Recommendation}. In \bibinfo{booktitle}{\emph{Proceedings of the AAAI Conference on Artificial Intelligence}}, Vol.~\bibinfo{volume}{39}. \bibinfo{pages}{11790--11798}.
\newblock


\bibitem[Hu et~al\mbox{.}(2024)]%
        {hu2024mgdcf}
\bibfield{author}{\bibinfo{person}{Jun Hu}, \bibinfo{person}{Bryan Hooi}, \bibinfo{person}{Shengsheng Qian}, \bibinfo{person}{Quan Fang}, {and} \bibinfo{person}{Changsheng Xu}.} \bibinfo{year}{2024}\natexlab{}.
\newblock \showarticletitle{MGDCF: Distance learning via Markov graph diffusion for neural collaborative filtering}.
\newblock \bibinfo{journal}{\emph{IEEE Transactions on Knowledge and Data Engineering}} \bibinfo{volume}{36}, \bibinfo{number}{7} (\bibinfo{year}{2024}), \bibinfo{pages}{3281--3296}.
\newblock


\bibitem[Jamali and Ester(2010)]%
        {jamali2010matrix}
\bibfield{author}{\bibinfo{person}{Mohsen Jamali} {and} \bibinfo{person}{Martin Ester}.} \bibinfo{year}{2010}\natexlab{}.
\newblock \showarticletitle{A matrix factorization technique with trust propagation for recommendation in social networks}. In \bibinfo{booktitle}{\emph{Recsys}}. \bibinfo{pages}{135--142}.
\newblock


\bibitem[J{\"a}rvelin and Kek{\"a}l{\"a}inen(2002)]%
        {jarvelin2002cumulated}
\bibfield{author}{\bibinfo{person}{Kalervo J{\"a}rvelin} {and} \bibinfo{person}{Jaana Kek{\"a}l{\"a}inen}.} \bibinfo{year}{2002}\natexlab{}.
\newblock \showarticletitle{Cumulated gain-based evaluation of IR techniques}.
\newblock \bibinfo{journal}{\emph{ACM Transactions on Information Systems (TOIS)}} \bibinfo{volume}{20}, \bibinfo{number}{4} (\bibinfo{year}{2002}), \bibinfo{pages}{422--446}.
\newblock


\bibitem[Jiang et~al\mbox{.}(2024)]%
        {jiang2024challenging}
\bibfield{author}{\bibinfo{person}{Wei Jiang}, \bibinfo{person}{Xinyi Gao}, \bibinfo{person}{Guandong Xu}, \bibinfo{person}{Tong Chen}, {and} \bibinfo{person}{Hongzhi Yin}.} \bibinfo{year}{2024}\natexlab{}.
\newblock \showarticletitle{Challenging Low Homophily in Social Recommendation}. In \bibinfo{booktitle}{\emph{TheWebConf}}. \bibinfo{pages}{3476--3484}.
\newblock


\bibitem[Kingma and Ba(2014)]%
        {kingma2014adam}
\bibfield{author}{\bibinfo{person}{Diederik~P Kingma} {and} \bibinfo{person}{Jimmy Ba}.} \bibinfo{year}{2014}\natexlab{}.
\newblock \showarticletitle{Adam: A method for stochastic optimization}.
\newblock \bibinfo{journal}{\emph{arXiv preprint arXiv:1412.6980}} (\bibinfo{year}{2014}).
\newblock


\bibitem[Konstas et~al\mbox{.}(2009)]%
        {konstas2009social}
\bibfield{author}{\bibinfo{person}{Ioannis Konstas}, \bibinfo{person}{Vassilios Stathopoulos}, {and} \bibinfo{person}{Joemon~M Jose}.} \bibinfo{year}{2009}\natexlab{}.
\newblock \showarticletitle{On social networks and collaborative recommendation}. In \bibinfo{booktitle}{\emph{SIGIR}}. \bibinfo{pages}{195--202}.
\newblock


\bibitem[Krueger et~al\mbox{.}(2021)]%
        {vRex}
\bibfield{author}{\bibinfo{person}{David Krueger}, \bibinfo{person}{Ethan Caballero}, \bibinfo{person}{Joern-Henrik Jacobsen}, \bibinfo{person}{Amy Zhang}, \bibinfo{person}{Jonathan Binas}, \bibinfo{person}{Dinghuai Zhang}, \bibinfo{person}{Remi Le~Priol}, {and} \bibinfo{person}{Aaron Courville}.} \bibinfo{year}{2021}\natexlab{}.
\newblock \showarticletitle{Out-of-distribution generalization via risk extrapolation (rex)}. In \bibinfo{booktitle}{\emph{ICML}}. PMLR, \bibinfo{pages}{5815--5826}.
\newblock


\bibitem[Li et~al\mbox{.}(2022)]%
        {li2022learning}
\bibfield{author}{\bibinfo{person}{Haoyang Li}, \bibinfo{person}{Ziwei Zhang}, \bibinfo{person}{Xin Wang}, {and} \bibinfo{person}{Wenwu Zhu}.} \bibinfo{year}{2022}\natexlab{}.
\newblock \showarticletitle{Learning invariant graph representations for out-of-distribution generalization}.
\newblock \bibinfo{journal}{\emph{NeurIPS}}  \bibinfo{volume}{35} (\bibinfo{year}{2022}), \bibinfo{pages}{11828--11841}.
\newblock


\bibitem[Liao et~al\mbox{.}(2022)]%
        {liao2022sociallgn}
\bibfield{author}{\bibinfo{person}{Jie Liao}, \bibinfo{person}{Wei Zhou}, \bibinfo{person}{Fengji Luo}, \bibinfo{person}{Junhao Wen}, \bibinfo{person}{Min Gao}, \bibinfo{person}{Xiuhua Li}, {and} \bibinfo{person}{Jun Zeng}.} \bibinfo{year}{2022}\natexlab{}.
\newblock \showarticletitle{SocialLGN: Light graph convolution network for social recommendation}.
\newblock \bibinfo{journal}{\emph{Information Sciences}}  \bibinfo{volume}{589} (\bibinfo{year}{2022}), \bibinfo{pages}{595--607}.
\newblock


\bibitem[Ma et~al\mbox{.}(2008)]%
        {ma2008sorec}
\bibfield{author}{\bibinfo{person}{Hao Ma}, \bibinfo{person}{Haixuan Yang}, \bibinfo{person}{Michael~R Lyu}, {and} \bibinfo{person}{Irwin King}.} \bibinfo{year}{2008}\natexlab{}.
\newblock \showarticletitle{Sorec: social recommendation using probabilistic matrix factorization}. In \bibinfo{booktitle}{\emph{CIKM}}. \bibinfo{pages}{931--940}.
\newblock


\bibitem[Ma et~al\mbox{.}(2024b)]%
        {ma2024multimodal}
\bibfield{author}{\bibinfo{person}{Haokai Ma}, \bibinfo{person}{Yimeng Yang}, \bibinfo{person}{Lei Meng}, \bibinfo{person}{Ruobing Xie}, {and} \bibinfo{person}{Xiangxu Meng}.} \bibinfo{year}{2024}\natexlab{b}.
\newblock \showarticletitle{Multimodal conditioned diffusion model for recommendation}. In \bibinfo{booktitle}{\emph{Companion Proceedings of the ACM Web Conference 2024}}. \bibinfo{pages}{1733--1740}.
\newblock


\bibitem[Ma et~al\mbox{.}(2011)]%
        {ma2011recommender}
\bibfield{author}{\bibinfo{person}{Hao Ma}, \bibinfo{person}{Dengyong Zhou}, \bibinfo{person}{Chao Liu}, \bibinfo{person}{Michael~R Lyu}, {and} \bibinfo{person}{Irwin King}.} \bibinfo{year}{2011}\natexlab{}.
\newblock \showarticletitle{Recommender systems with social regularization}. In \bibinfo{booktitle}{\emph{WSDM}}. \bibinfo{pages}{287--296}.
\newblock


\bibitem[Ma et~al\mbox{.}(2024a)]%
        {wsdm2024madm}
\bibfield{author}{\bibinfo{person}{Wenze Ma}, \bibinfo{person}{Yuexian Wang}, \bibinfo{person}{Yanmin Zhu}, \bibinfo{person}{Zhaobo Wang}, \bibinfo{person}{Mengyuan Jing}, \bibinfo{person}{Xuhao Zhao}, \bibinfo{person}{Jiadi Yu}, {and} \bibinfo{person}{Feilong Tang}.} \bibinfo{year}{2024}\natexlab{a}.
\newblock \showarticletitle{MADM: A Model-agnostic Denoising Module for Graph-based Social Recommendation}. In \bibinfo{booktitle}{\emph{WSDM}}. \bibinfo{pages}{501--509}.
\newblock


\bibitem[Marsden and Friedkin(1993)]%
        {marsden1993network}
\bibfield{author}{\bibinfo{person}{Peter~V Marsden} {and} \bibinfo{person}{Noah~E Friedkin}.} \bibinfo{year}{1993}\natexlab{}.
\newblock \showarticletitle{Network studies of social influence}.
\newblock \bibinfo{journal}{\emph{Sociological Methods \& Research}} \bibinfo{volume}{22}, \bibinfo{number}{1} (\bibinfo{year}{1993}), \bibinfo{pages}{127--151}.
\newblock


\bibitem[McPherson et~al\mbox{.}(2001)]%
        {mcpherson2001birds}
\bibfield{author}{\bibinfo{person}{Miller McPherson}, \bibinfo{person}{Lynn Smith-Lovin}, {and} \bibinfo{person}{James~M Cook}.} \bibinfo{year}{2001}\natexlab{}.
\newblock \showarticletitle{Birds of a feather: Homophily in social networks}.
\newblock \bibinfo{journal}{\emph{Annual review of sociology}} \bibinfo{volume}{27}, \bibinfo{number}{1} (\bibinfo{year}{2001}), \bibinfo{pages}{415--444}.
\newblock


\bibitem[Quan et~al\mbox{.}(2023)]%
        {WWW2023robust}
\bibfield{author}{\bibinfo{person}{Yuhan Quan}, \bibinfo{person}{Jingtao Ding}, \bibinfo{person}{Chen Gao}, \bibinfo{person}{Lingling Yi}, \bibinfo{person}{Depeng Jin}, {and} \bibinfo{person}{Yong Li}.} \bibinfo{year}{2023}\natexlab{}.
\newblock \showarticletitle{Robust Preference-Guided Denoising for Graph based Social Recommendation}. In \bibinfo{booktitle}{\emph{TheWebConf}}. \bibinfo{pages}{1097--1108}.
\newblock


\bibitem[Rendle et~al\mbox{.}(2009)]%
        {UAI2009BPR}
\bibfield{author}{\bibinfo{person}{Steffen Rendle}, \bibinfo{person}{Christoph Freudenthaler}, \bibinfo{person}{Zeno Gantner}, {and} \bibinfo{person}{Lars Schmidt-Thieme}.} \bibinfo{year}{2009}\natexlab{}.
\newblock \showarticletitle{BPR: Bayesian personalized ranking from implicit feedback}. In \bibinfo{booktitle}{\emph{UAI}}. \bibinfo{pages}{452--461}.
\newblock


\bibitem[Tang et~al\mbox{.}(2013)]%
        {tang2013social}
\bibfield{author}{\bibinfo{person}{Jiliang Tang}, \bibinfo{person}{Xia Hu}, {and} \bibinfo{person}{Huan Liu}.} \bibinfo{year}{2013}\natexlab{}.
\newblock \showarticletitle{Social recommendation: a review}.
\newblock \bibinfo{journal}{\emph{Social Network Analysis and Mining}}  \bibinfo{volume}{3} (\bibinfo{year}{2013}), \bibinfo{pages}{1113--1133}.
\newblock


\bibitem[Wang et~al\mbox{.}(2021b)]%
        {TOIS2021hypersorec}
\bibfield{author}{\bibinfo{person}{Hao Wang}, \bibinfo{person}{Defu Lian}, \bibinfo{person}{Hanghang Tong}, \bibinfo{person}{Qi Liu}, \bibinfo{person}{Zhenya Huang}, {and} \bibinfo{person}{Enhong Chen}.} \bibinfo{year}{2021}\natexlab{b}.
\newblock \showarticletitle{Hypersorec: Exploiting hyperbolic user and item representations with multiple aspects for social-aware recommendation}.
\newblock \bibinfo{journal}{\emph{TOIS}} \bibinfo{volume}{40}, \bibinfo{number}{2} (\bibinfo{year}{2021}), \bibinfo{pages}{1--28}.
\newblock


\bibitem[Wang et~al\mbox{.}(2024)]%
        {WWW2024unleashing}
\bibfield{author}{\bibinfo{person}{Shuyao Wang}, \bibinfo{person}{Yongduo Sui}, \bibinfo{person}{Chao Wang}, {and} \bibinfo{person}{Hui Xiong}.} \bibinfo{year}{2024}\natexlab{}.
\newblock \showarticletitle{Unleashing the Power of Knowledge Graph for Recommendation via Invariant Learning}. In \bibinfo{booktitle}{\emph{TheWebConf}}. \bibinfo{pages}{3745--3755}.
\newblock


\bibitem[Wang et~al\mbox{.}(2021a)]%
        {wang2021denoising}
\bibfield{author}{\bibinfo{person}{Wenjie Wang}, \bibinfo{person}{Fuli Feng}, \bibinfo{person}{Xiangnan He}, \bibinfo{person}{Liqiang Nie}, {and} \bibinfo{person}{Tat-Seng Chua}.} \bibinfo{year}{2021}\natexlab{a}.
\newblock \showarticletitle{Denoising implicit feedback for recommendation}. In \bibinfo{booktitle}{\emph{WSDM}}. \bibinfo{pages}{373--381}.
\newblock


\bibitem[Wang et~al\mbox{.}(2022)]%
        {wang2022invariant}
\bibfield{author}{\bibinfo{person}{Zimu Wang}, \bibinfo{person}{Yue He}, \bibinfo{person}{Jiashuo Liu}, \bibinfo{person}{Wenchao Zou}, \bibinfo{person}{Philip~S Yu}, {and} \bibinfo{person}{Peng Cui}.} \bibinfo{year}{2022}\natexlab{}.
\newblock \showarticletitle{Invariant preference learning for general debiasing in recommendation}. In \bibinfo{booktitle}{\emph{KDD}}. \bibinfo{pages}{1969--1978}.
\newblock


\bibitem[Wu et~al\mbox{.}(2024)]%
        {wu2024effectiveness}
\bibfield{author}{\bibinfo{person}{Jiancan Wu}, \bibinfo{person}{Xiang Wang}, \bibinfo{person}{Xingyu Gao}, \bibinfo{person}{Jiawei Chen}, \bibinfo{person}{Hongcheng Fu}, {and} \bibinfo{person}{Tianyu Qiu}.} \bibinfo{year}{2024}\natexlab{}.
\newblock \showarticletitle{On the effectiveness of sampled softmax loss for item recommendation}.
\newblock \bibinfo{journal}{\emph{TOIS}} \bibinfo{volume}{42}, \bibinfo{number}{4} (\bibinfo{year}{2024}), \bibinfo{pages}{1--26}.
\newblock


\bibitem[Wu et~al\mbox{.}(2020)]%
        {wu2020diffnet++}
\bibfield{author}{\bibinfo{person}{Le Wu}, \bibinfo{person}{Junwei Li}, \bibinfo{person}{Peijie Sun}, \bibinfo{person}{Richang Hong}, \bibinfo{person}{Yong Ge}, {and} \bibinfo{person}{Meng Wang}.} \bibinfo{year}{2020}\natexlab{}.
\newblock \showarticletitle{Diffnet++: A neural influence and interest diffusion network for social recommendation}.
\newblock \bibinfo{journal}{\emph{TKDE}} (\bibinfo{year}{2020}).
\newblock


\bibitem[Wu et~al\mbox{.}(2019)]%
        {DiffNet}
\bibfield{author}{\bibinfo{person}{Le Wu}, \bibinfo{person}{Peijie Sun}, \bibinfo{person}{Yanjie Fu}, \bibinfo{person}{Richang Hong}, \bibinfo{person}{Xiting Wang}, {and} \bibinfo{person}{Meng Wang}.} \bibinfo{year}{2019}\natexlab{}.
\newblock \showarticletitle{A neural influence diffusion model for social recommendation}. In \bibinfo{booktitle}{\emph{SIGIR}}. \bibinfo{pages}{235--244}.
\newblock


\bibitem[Wu et~al\mbox{.}(2022b)]%
        {wu2022handling}
\bibfield{author}{\bibinfo{person}{Qitian Wu}, \bibinfo{person}{Hengrui Zhang}, \bibinfo{person}{Junchi Yan}, {and} \bibinfo{person}{David Wipf}.} \bibinfo{year}{2022}\natexlab{b}.
\newblock \showarticletitle{Handling distribution shifts on graphs: An invariance perspective}.
\newblock \bibinfo{journal}{\emph{ICLR}} (\bibinfo{year}{2022}).
\newblock


\bibitem[Wu et~al\mbox{.}(2022a)]%
        {ICLR2022discovering}
\bibfield{author}{\bibinfo{person}{Ying-Xin Wu}, \bibinfo{person}{Xiang Wang}, \bibinfo{person}{An Zhang}, \bibinfo{person}{Xiangnan He}, {and} \bibinfo{person}{Tat-Seng Chua}.} \bibinfo{year}{2022}\natexlab{a}.
\newblock \showarticletitle{Discovering invariant rationales for graph neural networks}.
\newblock \bibinfo{journal}{\emph{ICLR}} (\bibinfo{year}{2022}).
\newblock


\bibitem[Xia et~al\mbox{.}(2024)]%
        {xia2024learning}
\bibfield{author}{\bibinfo{person}{Donglin Xia}, \bibinfo{person}{Xiao Wang}, \bibinfo{person}{Nian Liu}, {and} \bibinfo{person}{Chuan Shi}.} \bibinfo{year}{2024}\natexlab{}.
\newblock \showarticletitle{Learning invariant representations of graph neural networks via cluster generalization}.
\newblock \bibinfo{journal}{\emph{NeurIPS}}  \bibinfo{volume}{36} (\bibinfo{year}{2024}).
\newblock


\bibitem[Yang et~al\mbox{.}(2025)]%
        {yang2025less}
\bibfield{author}{\bibinfo{person}{Yonghui Yang}, \bibinfo{person}{Le Wu}, \bibinfo{person}{Zhuangzhuang He}, \bibinfo{person}{Zhengwei Wu}, \bibinfo{person}{Richang Hong}, {and} \bibinfo{person}{Meng Wang}.} \bibinfo{year}{2025}\natexlab{}.
\newblock \showarticletitle{Less is More: Information Bottleneck Denoised Multimedia Recommendation}.
\newblock \bibinfo{journal}{\emph{arXiv preprint arXiv:2501.12175}} (\bibinfo{year}{2025}).
\newblock


\bibitem[Yang et~al\mbox{.}(2021)]%
        {yang2021enhanced}
\bibfield{author}{\bibinfo{person}{Yonghui Yang}, \bibinfo{person}{Le Wu}, \bibinfo{person}{Richang Hong}, \bibinfo{person}{Kun Zhang}, {and} \bibinfo{person}{Meng Wang}.} \bibinfo{year}{2021}\natexlab{}.
\newblock \showarticletitle{Enhanced graph learning for collaborative filtering via mutual information maximization}. In \bibinfo{booktitle}{\emph{SIGIR}}. \bibinfo{pages}{71--80}.
\newblock


\bibitem[Yang et~al\mbox{.}(2024)]%
        {yang2024graph}
\bibfield{author}{\bibinfo{person}{Yonghui Yang}, \bibinfo{person}{Le Wu}, \bibinfo{person}{Zihan Wang}, \bibinfo{person}{Zhuangzhuang He}, \bibinfo{person}{Richang Hong}, {and} \bibinfo{person}{Meng Wang}.} \bibinfo{year}{2024}\natexlab{}.
\newblock \showarticletitle{Graph Bottlenecked Social Recommendation}.
\newblock \bibinfo{journal}{\emph{KDD}} (\bibinfo{year}{2024}).
\newblock


\bibitem[Yang et~al\mbox{.}(2023b)]%
        {HGSR}
\bibfield{author}{\bibinfo{person}{Yonghui Yang}, \bibinfo{person}{Le Wu}, \bibinfo{person}{Kun Zhang}, \bibinfo{person}{Richang Hong}, \bibinfo{person}{Hailin Zhou}, \bibinfo{person}{Zhiqiang Zhang}, \bibinfo{person}{Jun Zhou}, {and} \bibinfo{person}{Meng Wang}.} \bibinfo{year}{2023}\natexlab{b}.
\newblock \showarticletitle{Hyperbolic Graph Learning for Social Recommendation}.
\newblock \bibinfo{journal}{\emph{IEEE TKDE}} (\bibinfo{year}{2023}).
\newblock


\bibitem[Yang et~al\mbox{.}(2023a)]%
        {yang2023generative}
\bibfield{author}{\bibinfo{person}{Yonghui Yang}, \bibinfo{person}{Zhengwei Wu}, \bibinfo{person}{Le Wu}, \bibinfo{person}{Kun Zhang}, \bibinfo{person}{Richang Hong}, \bibinfo{person}{Zhiqiang Zhang}, \bibinfo{person}{Jun Zhou}, {and} \bibinfo{person}{Meng Wang}.} \bibinfo{year}{2023}\natexlab{a}.
\newblock \showarticletitle{Generative-contrastive graph learning for recommendation}. In \bibinfo{booktitle}{\emph{SIGIR}}. \bibinfo{pages}{1117--1126}.
\newblock


\bibitem[Yu et~al\mbox{.}(2021a)]%
        {SEPT}
\bibfield{author}{\bibinfo{person}{Junliang Yu}, \bibinfo{person}{Hongzhi Yin}, \bibinfo{person}{Min Gao}, \bibinfo{person}{Xin Xia}, \bibinfo{person}{Xiangliang Zhang}, {and} \bibinfo{person}{Nguyen~Quoc Viet~Hung}.} \bibinfo{year}{2021}\natexlab{a}.
\newblock \showarticletitle{Socially-aware self-supervised tri-training for recommendation}. In \bibinfo{booktitle}{\emph{SIGKDD}}. \bibinfo{pages}{2084--2092}.
\newblock


\bibitem[Yu et~al\mbox{.}(2020)]%
        {TKDE2020enhancing}
\bibfield{author}{\bibinfo{person}{Junliang Yu}, \bibinfo{person}{Hongzhi Yin}, \bibinfo{person}{Jundong Li}, \bibinfo{person}{Min Gao}, \bibinfo{person}{Zi Huang}, {and} \bibinfo{person}{Lizhen Cui}.} \bibinfo{year}{2020}\natexlab{}.
\newblock \showarticletitle{Enhancing social recommendation with adversarial graph convolutional networks}.
\newblock \bibinfo{journal}{\emph{IEEE TKDE}} \bibinfo{volume}{34}, \bibinfo{number}{8} (\bibinfo{year}{2020}), \bibinfo{pages}{3727--3739}.
\newblock


\bibitem[Yu et~al\mbox{.}(2021b)]%
        {yu2021self}
\bibfield{author}{\bibinfo{person}{Junliang Yu}, \bibinfo{person}{Hongzhi Yin}, \bibinfo{person}{Jundong Li}, \bibinfo{person}{Qinyong Wang}, \bibinfo{person}{Nguyen Quoc~Viet Hung}, {and} \bibinfo{person}{Xiangliang Zhang}.} \bibinfo{year}{2021}\natexlab{b}.
\newblock \showarticletitle{Self-supervised multi-channel hypergraph convolutional network for social recommendation}. In \bibinfo{booktitle}{\emph{TheWebConf}}. \bibinfo{pages}{413--424}.
\newblock


\bibitem[Yue et~al\mbox{.}(2025)]%
        {yue2025learning}
\bibfield{author}{\bibinfo{person}{Linan Yue}, \bibinfo{person}{Qi Liu}, \bibinfo{person}{Ye Liu}, \bibinfo{person}{Weibo Gao}, {and} \bibinfo{person}{Fangzhou Yao}.} \bibinfo{year}{2025}\natexlab{}.
\newblock \showarticletitle{Learning from shortcut: a shortcut-guided approach for explainable graph learning}.
\newblock \bibinfo{journal}{\emph{Frontiers of Computer Science}} \bibinfo{volume}{19}, \bibinfo{number}{8} (\bibinfo{year}{2025}), \bibinfo{pages}{198338}.
\newblock


\bibitem[Zhang et~al\mbox{.}(2023)]%
        {WWW2023invariant}
\bibfield{author}{\bibinfo{person}{An Zhang}, \bibinfo{person}{Jingnan Zheng}, \bibinfo{person}{Xiang Wang}, \bibinfo{person}{Yancheng Yuan}, {and} \bibinfo{person}{Tat-Seng Chua}.} \bibinfo{year}{2023}\natexlab{}.
\newblock \showarticletitle{Invariant collaborative filtering to popularity distribution shift}. In \bibinfo{booktitle}{\emph{TheWebConf}}. \bibinfo{pages}{1240--1251}.
\newblock


\bibitem[Zhao et~al\mbox{.}(2020)]%
        {zhao2020revisiting}
\bibfield{author}{\bibinfo{person}{Wayne~Xin Zhao}, \bibinfo{person}{Junhua Chen}, \bibinfo{person}{Pengfei Wang}, \bibinfo{person}{Qi Gu}, {and} \bibinfo{person}{Ji-Rong Wen}.} \bibinfo{year}{2020}\natexlab{}.
\newblock \showarticletitle{Revisiting alternative experimental settings for evaluating top-n item recommendation algorithms}. In \bibinfo{booktitle}{\emph{CIKM}}. \bibinfo{pages}{2329--2332}.
\newblock


\end{thebibliography}
\end{document}